# Time-domain response of the ARIANNA detector

S.W. Barwick [a], E.C. Berg [a], D.Z. Besson [b,f], T. Duffin [a], J.C. Hanson [a,b,*], S.R. Klein [d], S.A. Kleinfelder [c], M. Piasecki [e], K. Ratzlaff [e], C. Reed [a], M. Roumi [c], T. Stezelberger [d], J. Tatar [a,g], J. Walker [a], R. Young [e], L. Zou [c]

[a] Dept. of Physics and Astronomy, University of California, Irvine, United States
[b] Dept. of Physics and Astronomy, University of Kansas, United States
[c] Dept. of Electrical Engineering and Computer Science, University of California, Irvine, United States
[d] Lawrence Berkeley National Laboratory, United States
[e] Instrumentation Design Lab, University of Kansas, United States
[f] Moscow Physics and Engineering Institute, United States
[g] Dept. of Physics, University of Washington, United States

ABSTRACT

The Antarctic Ross Ice Shelf Antenna Neutrino Array (ARIANNA) is a high-energy neutrino detector designed to record the Askaryan electric field signature of cosmogenic neutrino interactions in ice. To understand the inherent radio-frequency (RF) neutrino signature, the time-domain response of the ARIANNA RF receiver must be measured. ARIANNA uses Create CLP5130-2N log-periodic dipole arrays (LPDAs). The associated *effective height* operator converts incident electric fields to voltage waveforms at the LDPA terminals. The effective height versus time and incident angle was measured, along with the associated response of the ARIANNA RF amplifier. The results are verified by correlating to field measurements in air and ice, using oscilloscopes. Finally, theoretical models for the Askaryan electric field are combined with the detector response to predict the neutrino signature.

## 1. Introduction and theory

The origin of ultra-high energy cosmic rays (UHECRs) is an enduring mystery in astrophysics. Because cosmic-rays with energies above $4 \times 10^{19}$ eV interact with the cosmic microwave background radiation, they have a limited range, so terrestrial cosmic-ray detectors can only probe UHECR sources within $\approx 75$ Mpc. In these interactions, cosmic-ray protons are excited to a $\Delta^+$ resonance; when the $\Delta^+$ decays, it produces neutrinos with energies in the range of $10^{17}$–$10^{20}$ eV [1–3]. These neutrinos can be used to probe UHECR sources at cosmic distances.

ARIANNA is designed to detect these neutrinos [4], by observing the coherent radio Cerenkov emission produced when UHE neutrinos interact in Antarctic ice. ARIANNA is located in Moore's Bay on the Ross Ice Shelf, about 100 miles south of McMurdo station. There, the ice is about 570 m thick [5,6], with the Ross Sea beneath it. ARIANNA is placed in Moore's Bay because aerial radar surveys show that the ice-water interface there is smooth and undisturbed, enabling clean reflections of downward-going radio waves off of the interface [7].

High energy particle cascades in ice contain an excess of negatively charged particles (mostly electrons), and radiates coherently at wavelengths that are large compared with the transverse size of the cascade [8]. For a cascade containing $10^{11}$ particles, the RF radiation is greatly enhanced. When observed along the Cerenkov cone, the radio signal grows linearly with frequency, up to a cutoff given by the size of the negative charge excess – about 1 GHz in ice. Away from the cone, the cutoff frequency is lower; one can use the observed spectrum to determine how far the detector is from the Cerenkov cone.

A number of experiments have used the Askaryan effect to search for cosmogenic neutrinos. One of the first was RICE, which placed antennas in holes drilled at the South Pole [9]. It pioneered the concept of using Antartic ice to search for neutrinos via radio waves. The ANITA experiment flies an array of horn antennas in the skies above Antarctica, to search for neutrino interactions in the ice [10], having completed two missions thus far. A number of experiments have also used radio-telescopes to search for ultra-high energy ($> 10^{20}$ eV) neutrino interaction in the Moon [11]. Detectors under construction, like ARIANNA (Moore's Bay) and ARA (South Pole), seek to observe the GZK neutrino flux via surface-array detectors.

* Corresponding author at: University of Kansas, United States.
*E-mail address:* j529h838@ku.edu (J.C. Hanson).

ARIANNA detects the radio waves using antennas that are buried shallowly in the ice. The current prototype stations each have four log-periodic dipole antennas (LPDAs) that are buried in a square pattern, facing downward. The antennas on opposite sides of the square are separated by 6 m. The direction to the neutrino interaction can be found by cross-correlating signals in these opposite antennas. The signal polarization is measured by comparing the signals from adjacent antennas. Between this polarization measurement and the determination of the frequency spectrum, the neutrino arrival direction can be determined, given enough channels above background [12] (doubling the channel number from 4 to 8 improves the solution). The amplitude of the waveforms, corrected for distance and ice absorption, provide knowledge of the interaction energy. The shape of the initial electric field, derived from the observed waveforms, would point to variables like the angle with respect to Cerenkov angle and the hadronic/electromagnetic nature of the event.

Monte Carlo simulations have been used to predict the signals that would be produced by neutrino interactions in the ice [13,14]. Both charged current (CC) and neutral current (NC) interactions are of interest. In NC interactions, neutrinos scatter from a target nucleus, depositing an average of 20% of their energy in the ice, in the form of a hadronic cascade. Charged-current interactions are similar, except that the neutrino produces a lepton which carries the remaining energy. Charged-current $\nu_e$ interactions produce an electron, creating an additional electromagnetic cascade (this discussion does not distinguish particle from anti-particle). Tau leptons from CC interactions also produce Askaryan pulses, albeit at a significant separation ($\approx$ 1 km) from the point of the neutrino interaction.

## 2. Antenna effective height

The antenna response to time-varying Askaryan electric field may be parameterized in terms of an effective height operator. This height may be determined by measuring the antenna response to a broad-band signal. To measure the LPDA effective height, two identical LPDAs were placed inside an anechoic chamber, facing each other. One LPDA transmitted an impulse, and the other received it, a distance $r$ away. The received signal voltage is shown in Eq. (1) and derived below.

$$V_L(t) = \frac{2^2}{2\pi rc}\left(\frac{Z_L Z_0}{(Z_L+Z_{in})^2}\right)\vec{h}_{rx}\circ\vec{h}_{rx}\circ\dot{V}_{src}(t) \quad (1)$$

In Eq. (1), the circles denote the convolution operator, $Z_0$ is the impedance of free space ($\approx 120\pi$), $Z_{in}$ is the antenna impedance ($\approx 50\Omega$), $c$ is the speed of light ($\approx$ 0.3 m/ns), $V_L$ is the received voltage at the antenna port, $V_{src}$ is the original voltage impulse transmitted through the first antenna, and $\vec{h}_{rx}$ is the vector-like effective height. Measurements in this work were co-polarized, so the vector symbols will be dropped from now on. For similar measurements of LPDA effective height characterization, see [15,16]. Convolution is linear, commutative, and overall time-invariant.

Eq. (1) is now derived. A receiver with input line impedance $Z_L$ is connected to an antenna of impedance $Z_{in}$. The voltage seen by the receiver is reduced by a factor

$$\frac{V_L}{V_{O.C.}} = \frac{Z_L}{Z_L+Z_{in}} \quad (2)$$

The open-circuit voltage of the antenna-receiver system is $V_{O.C.}$. The effective height is proportional to this factor [17,18]. The authors of [18] also include a factor of 2 such that the voltage delivered to the receiver is

$$V_L(t) = 2\left(\frac{Z_L}{Z_L+Z_{in}}\right)h_{rx}(t)\circ E(t) \quad (3)$$

The factor of two out front in Eq. (5) is consistent with [18], in which the effective height operator was used to predict (correctly) the Askaryan amplitude of an electromagnetic cascade in the lab. The factor of two corrects $V_L$ from the observed data on the 50Ω-scope to the voltage actually produced by the antenna. For the transmitter, the ratio of the voltage delivered to the antenna to the open-circuit voltage is [17]

$$\frac{V_A}{V_{O.C.}} = \left(\frac{Z_{in}}{Z_{in}+Z_L}\right)\left(\frac{Z_0}{Z_{in}}\right) \quad (4)$$

The additional factor $Z_0/Z_{in}$ is included to account for the antenna coupling the radiated electric field to the impedance of free space. The electric field radiated by the transmitter is therefore

$$E(t) = \frac{1}{2\pi rc}\left(\frac{Z_{in}}{Z_{in}+Z_L}\right)\left(\frac{Z_0}{Z_{in}}\right)h_{tx}\circ V_{src}(t) \quad (5)$$

Before combining Eqs. (3) and (5), the relationship between $h_{rx}$ and $h_{tx}$ must be mentioned, and it is shown in Eq. (6). The authors of [17,16] differ on the factor of 2 in Eq. (6), but only due to slightly different definitions of antenna quantities. One motivation of the time-derivative in Eq. (6) is that antennas do not radiate DC voltages; in the Fourier domain, time-derivation becomes multiplication of the original function by the frequency, so that $h_{rx}(\omega)\to 0$ for $\omega\to 0$.

$$h_{tx} = 2\partial_t h_{rx} \quad (6)$$

Integration by parts, and the *stability condition* of the linear time-invariant system ($h_{rx}(t)\to 0$ for $t\to\pm\infty$) allows the transfer of the time-derivative to the input signal. Therefore, Eq. (5) can be rewritten:

$$E(t) = \frac{1}{\pi rc}\left(\frac{Z_{in}}{Z_{in}+Z_L}\right)\left(\frac{Z_0}{Z_{in}}\right)h_{rx}\circ\dot{V}_{src}(t) \quad (7)$$

Combining (3) and (7) produces

$$V_L(t) = \frac{2^2}{2\pi rc}\left(\frac{Z_L Z_0}{(Z_L+Z_{in})^2}\right)h_{rx}\circ h_{rx}\circ\dot{V}_{src}(t) \quad (8)$$

The LPDAs in the anechoic chamber are co-polarized and impedance matched to the 50Ω input impedance of the oscilloscope and amplifier ($Z_L\approx Z_{in}$), so (8) simplifies to

$$V_L(t) = \frac{1}{2\pi rc}\left(\frac{Z_0}{Z_L}\right)h_{rx}\circ h_{rx}\circ\dot{V}_{src}(t) \quad (9)$$

Eq. (9) provides the basis for measuring the effective height. Using the effective height, and similar *transfer functions* for the other components in the ARIANNA data acquisition, neutrino signals may be predicted for a given model of Askaryan radiation. The impedance matching criterion is discussed further in the appendix.

## 3. Experimental technique

Fig. 1 contains a diagram of the LPDA, a Create CLP5130-2N, and defines the terms E- and H-plane, with respect to the antenna design. The geometry of the LPDA provides the advantage of frequency-independent broadband response, similar to a radio horn. The advantage over a horn (as is revealed by the data) is the uniformity of the response over a wide range of angles. The beam width, VSWR, radiated power, and gain are all favorable over the 105–1300 MHz range. Monte Carlo simulations [12] show that such an antenna maximizes broadband neutrino signal over backgrounds.

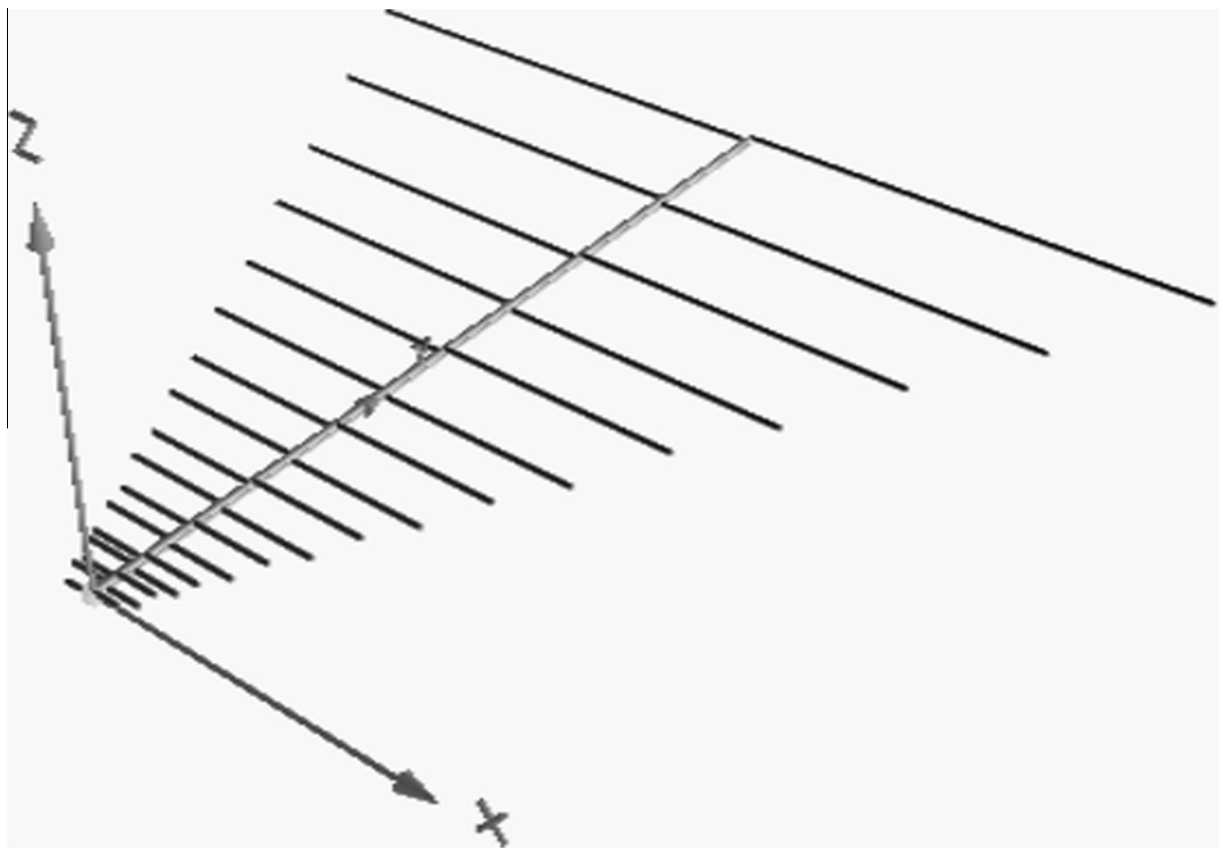

**Fig. 1.** A diagram of the LPDA, demonstrating the logarithmic spacing and dipole length. The *x*–*y* plane contains the dipole elements (*tines*), and is denoted the E-plane. The H-plane, in this case, is the *y*–*z* plane. The -*y* direction is the forward direction of the LPDA. The longest ($\lambda/2$) dipole is 1.45 m, and the length ratio of adjacent dipoles is 0.83. The 1.385 m-long *spine* holds the dipoles in place, and the feed point is at the shortest dipole (at the origin).

Two LPDAs were arranged a distance $r = 5.72$ m apart, facing towards each other (boresight configuration), inside of an anechoic chamber. A 400 ps wide impulse from an Avtech pulser was sent through coaxial cable to the transmitter, and the received signal was recorded on a Tektronix TDS5104 oscilloscope with a nominal bandwidth of 1 GHz. The spectrum of this raw impulse signal, as measured with the oscilloscope, matches the spectrum taken with a 2.5 GHz bandwidth spectrum analyzer up to a frequency of 1.25 GHz. Above this frequency, the oscilloscope begins to attenuate the data. Results above this benchmark frequency should not be trusted. Fig. 2 shows the experimental setup. The receiving LPDA was attached to a post via a rotating flange, and the post is fixed to a rotating turntable. The flange and turntable apparatus enabled independent rotation of the receiver in both $\theta$ and $\phi$.

The total system transfer function besides that of the antennas was shown to be a small correction, by comparing the raw pulse with the pulse propagated through all cables in the system (Fig. 3). The 400 ps width of the pulse sent to the transmitter was sufficiently narrow to guarantee that the highest frequencies of the LPDA were probed. To ensure that the entire bandwidth was accessible to the instrumentation, the voltage standing wave ratio (VSWR) was measured by a network analyzer built into the chamber systems (Fig. 3). The VSWR is a standard RF probe of antenna efficiency; a VSWR of 1.0 indicates all transmitted energy is being radiated with high efficiency.

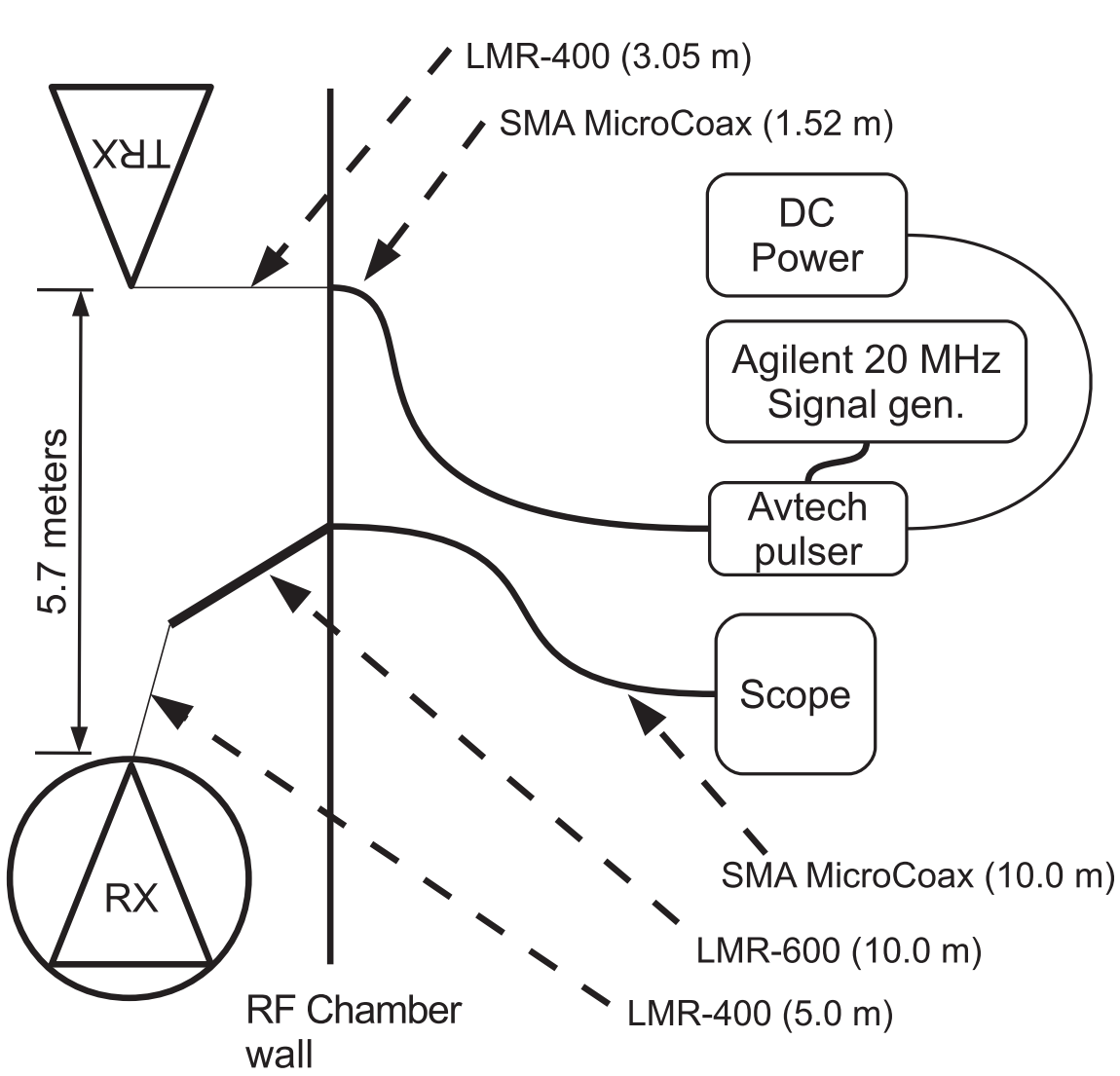

**Fig. 2.** The experimental setup of the bore-sight configuration. The Avtech was a model AVP-AV-1S type pulser, capable of 0.4 ns pulse widths.

In the anechoic chamber, the antenna was properly grounded and isolated from interfering electronic equipment, and the VSWR is near 1.0 for the experimental bandwidth. The VSWR of these antennas was measured previously during the installation of the first ARIANNA prototype station [19]. Those measurements recorded a VSWR of ≈1.5 from 100 MHz to above 1 GHz, when the antenna was ≈1.5 m above the snow surface. When the antenna was buried in densely packed snow, the VSWR was 1.5 down to 80 MHz. The index of refraction of the surface snow in Moore's Bay is $n = 1.29 \pm 0.02$ [5], and this result has been measured with several techniques [19,20]. The LPDA responds to the longest wavelength that physically can exist on the longest $\lambda/2$ dipoles. In a dielectric medium, the wavelength corresponding to a given frequency decreases by one factor of the index, meaning that 80 MHz can fit onto the 1.45 m LPDA dipoles (the longest pair). This effect is confirmed in NEC4 simulations [21], where the accompanying shift in antenna impedance is not enough to reduce the efficiency. This point will be discussed further in Section 5.2.

The programmable turntable was used to rotate the receiving antenna in both the E- and H-planes. The E-plane is the plane containing the LPDA tines (associated with the spherical coordinate $\phi$), and the H-plane (associated with the spherical coordinate $\theta$) is orthogonal to the E-plane, and contains the LPDA central spine (Fig. 1). For the pulsed measurements, waveforms were measured in 10 degree increments in the E and H-planes, beginning with 0° in each. The configuration at 0° in both planes is the boresight configuration, with the LPDA, receiver facing the transmitter head-on. The E-plane was fully probed by the pulsed data, and 67% of the H-plane was probed. The missing third of the H-plane arose from limitations in the rotating flange, which was not able to tilt beyond ±63° in the H-plane.

The response is not expected to vary significantly versus $\theta$ until the incident wave is outside the H-plane front lobe (Fig. 4). The network analyzer was used to map the radiation pattern, shown in Fig. 4. A simple model of log-periodic antennas [22] is shown as well. The model is tuned to have the same geometric parameters as the CLP5130-2N, and the same bandwidth and number of dipole elements. The agreement is good for the forward lobes in the E and H-planes, and the back lobe in the H-plane. The E-plane result agrees in overall scale, but fluctuates around the average model prediction. For the experimental bandwidth, a front-to-back ratio of $\approx -15$ dB is observed across the bandwidth, in accordance with the manufacturer specifications. The front-to-back ratio is the ratio of received power at 180° to 0° in the E-plane, for identical transmission signals. Note that, in Fig. 4, the H-plane angle of 0° indicates the forward direction. In subsequent figures, the H-plane is associated with the polar angle, $\theta$, in which case the forward direction is $\theta = 90°$, where the antenna rests in the *x*–*y* plane.

## 4. Results and analysis

### 4.1. Data

The data from the boresight configuration is shown in Fig. 5. The scope recorded a downward-chirping signal approximately 40 ns wide, probing all frequencies accessible to the LPDA, according to the VSWR measurements (Fig. 3). A reflection is observed 67.5 ns after the initial pulse, and can be verified with auto-correlation of the data. The analysis was performed only on the first 67.5 ns of data, before the reflection. Similar pulse data was taken at all available turntable angles, and the input pulse was not changed. The chirping form of the impulse data has a simple explanation: the placement of the feed point on the LDPA causes the radiated phase, and group delay, to have predictable dependencies on frequency:

$$\phi(\omega) = \frac{\pi}{\ln\tau}\ln\left(\frac{\omega}{\omega_1}\right) \tag{10}$$

$$\tau_g(\omega) \equiv -\frac{d\phi}{d\omega} = -(2\nu\ln\tau)^{-1} \tag{11}$$

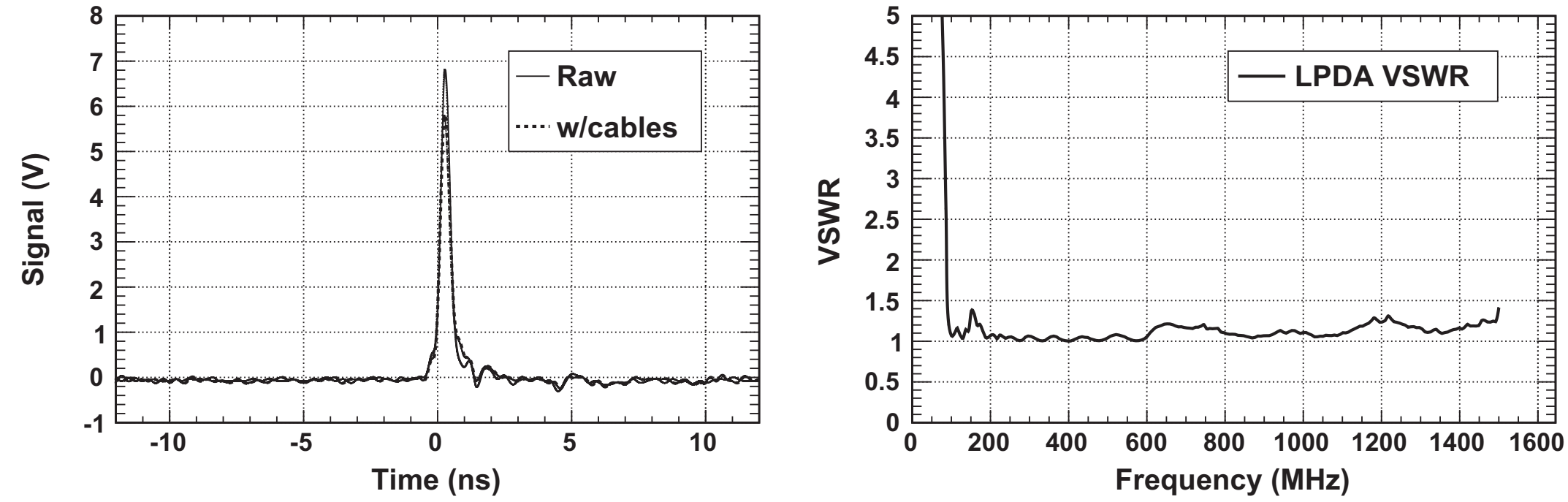

**Fig. 3.** (Left) The raw pulse sent to the transmitting antenna, as measured on the 1 GHz scope (Fig. 2), along with the raw pulse after having propagated through all coaxial cables in the system. (Right) The VSWR of the receiving antenna, measured by a network analyzer just outside the anechoic chamber.

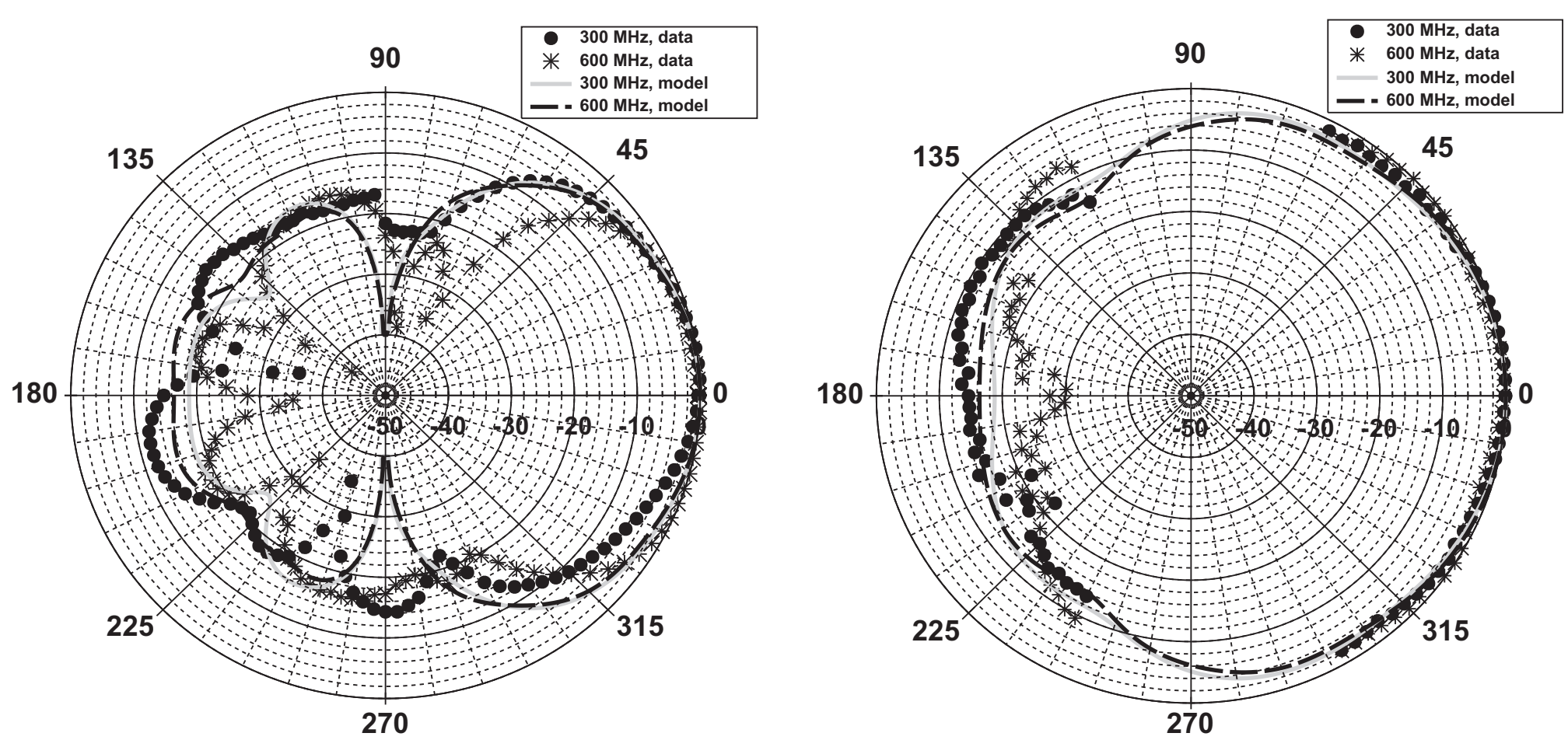

**Fig. 4.** The radiation pattern of the CLP-5130-2N LDPA (normalized to 0 dB maximum). The angles are relative to boresight in the counter-clockwise sense. (Left) the E-plane radiation pattern at 300 and 600 MHz, compared to a log-periodic simulation of the E-plane. (Right) the H-plane radiation pattern at 300 and 600 MHz, compared to the same simulation for H-plane. Note that 0° in the H-plane indicates the forward direction in this graph.

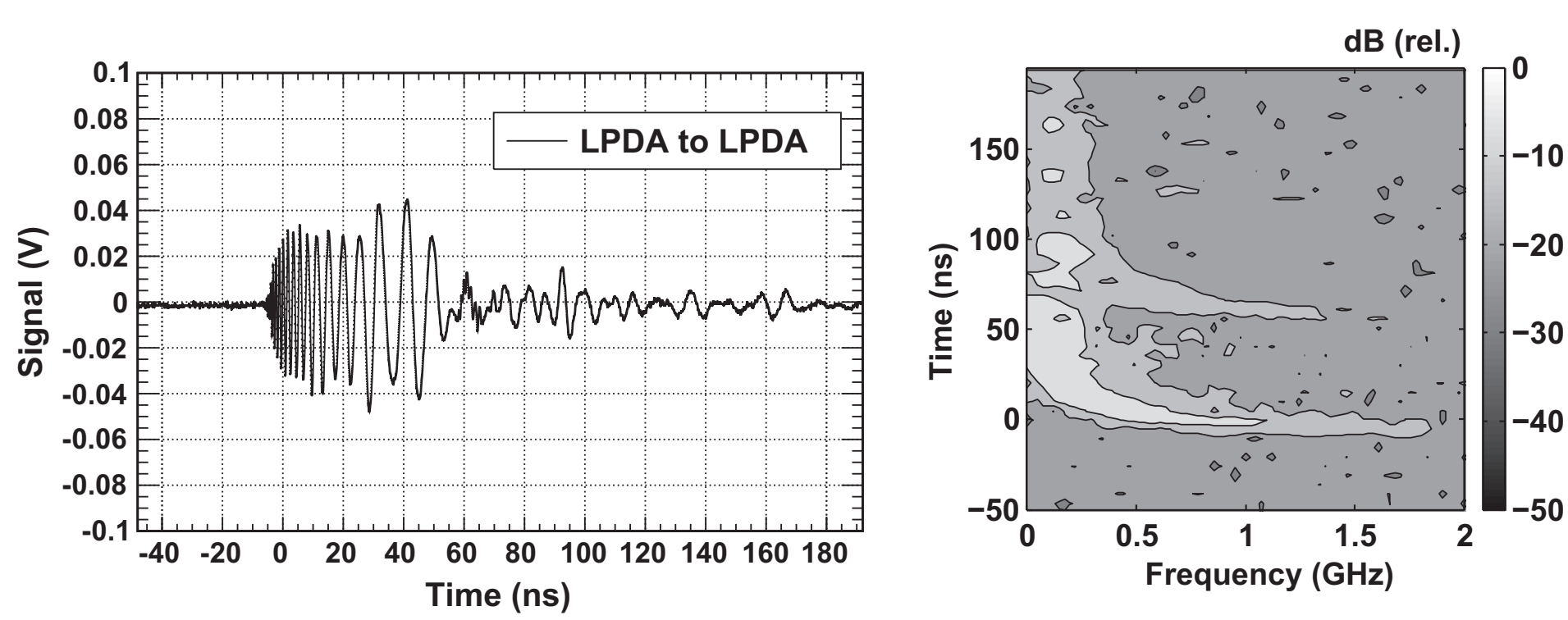

**Fig. 5.** (Left): The boresight time-domain results, shifted such that the trigger location is at 0 ns. (Right): a spectrogram of the boresight data, with the oscilloscope DC offset removed. The power is in units of dB, relative to maximum power.

The overall minus sign in the group delay $\tau_g(\omega)$ simply indicates a physical delay, rather than an earlier signal. In Eq. (10), $\omega_1$ is the angular frequency of the shortest radiating dipole, and $\tau$ is the ratio of the length of adjacent dipole elements. The location of the LPDA feed-point delays radiation of components with lower frequencies. Fortunately, this effect can be undone with a knowledge of the components' frequencies, and the LPDA $\tau$-parameter ($\tau = 0.83$ for the CLP-5130-2N) [23,24]. Because of the simple form of Eq. (10), an operator which commutes with the detector response can be constructed that removes the phase dispersion.

Examples of data from off-boresight configurations are shown in Figs. 6 and 7. These data demonstrate that it is not sufficient, for a broadband antenna, to model the time-response as the boresight function times the relative gain versus angle of the antenna. The radiation pattern depends on frequency, and the phasing must be studied at all angles to accurately predict signals. For example, a high-frequency incident plane wave interacting with the antenna at oblique angles changes the order in which antenna elements produce a voltage. Thus, the relative phasing between elements must depend on the orientation of the system.

### 4.2. Analysis

Eq. (9) represents the model used to explain the time-domain data. Eq. (9) can be re-written: $xV_L(t) - h \circ h \circ \dot{V}_{src}(t) = 0$, with $x = (Z_L/Z_0)2\pi rc$. An algorithm was developed to derive the waveform representation of h that solves this equation. Fig. 8 demonstrates how, after many iterations, a solution for $h$ was found that satisfies the equation with $x$ defined as above. Because of the shape of the LPDA, the expectation for the solution is a rapidly oscillating *chirp* that decreases in frequency as time increases.

During each iteration, the algorithm adds a small amount of white noise to the effective height samples $[s_1 : s_n]$, and nothing to subsequent samples $[s_{n+1} : s_N]$, where there are $N$ samples total. The result is kept only if the solution to $xV_L(t) - h \circ h \circ \dot{V}_{src}(t) = 0$ improves. Improvement is defined as a decrease in the mean squared-difference between model and data, denoted the *score*, S:

$$S = N^{-1}\sum_{i=1}^{N}\left(s_{i,data} - s_{i,model}\right)^2 \tag{12}$$

If the score decreases, there is a 1%-percent chance that $n \rightarrow n+1$ (a randomly drawn number between 0 and 1 must be less than 0.01). Thus, the algorithm focuses early action on the high-frequency content of the response, leaving the simpler low-frequency oscillations for the end of the calculation. The calculation terminates when the score has decreased by a factor of $\approx 100$, which is typical of convergence (subsequent iterations produce only marginal improvement).

The algorithm was applied to all the angles measured with the anechoic chamber turntable. The main lobe results converge and provide the effective height versus time. The LPDA null zones are not studied in detail, as the neutrino signals are heavily attenuated there. Several instances of the results are shown below in Fig. 9. The beam-width of the LPDA is 60° in the E-plane, meaning radiated power is reduced by 3 dB at ±30° from boresight. As the observation in the E-plane (Fig. 9, top) begins to take place outside the beam width, the amplitude shifts more rapidly, and is concentrated at lower (100–200 MHz) frequencies. The H-plane dependence (Fig. 9, bottom) is gradual, because the beam width is wider and more consistent in frequency.

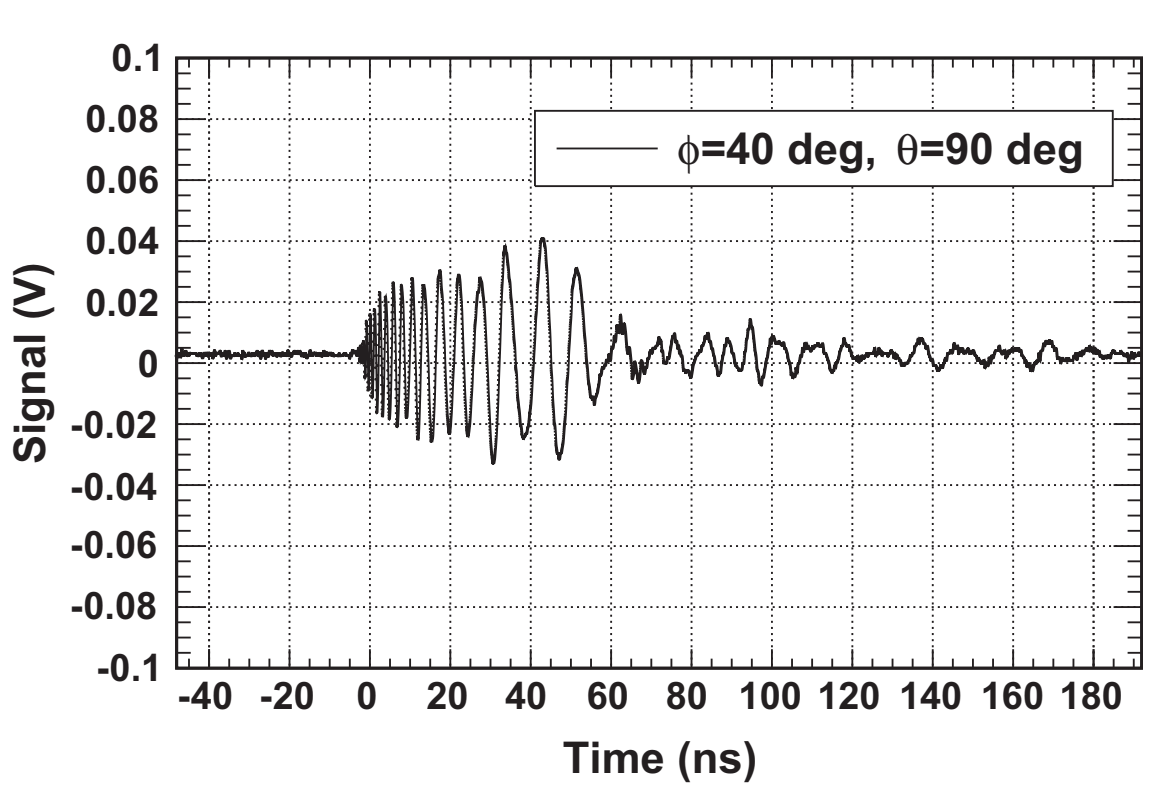

**Fig. 6.** The pulse data from the anechoic chamber at an angle of $\phi = 40°$ ($\phi$: angle in E-plane), and $\theta = 90°$ ($\theta$: angle in H-plane, with 90° corresponding to boresight).

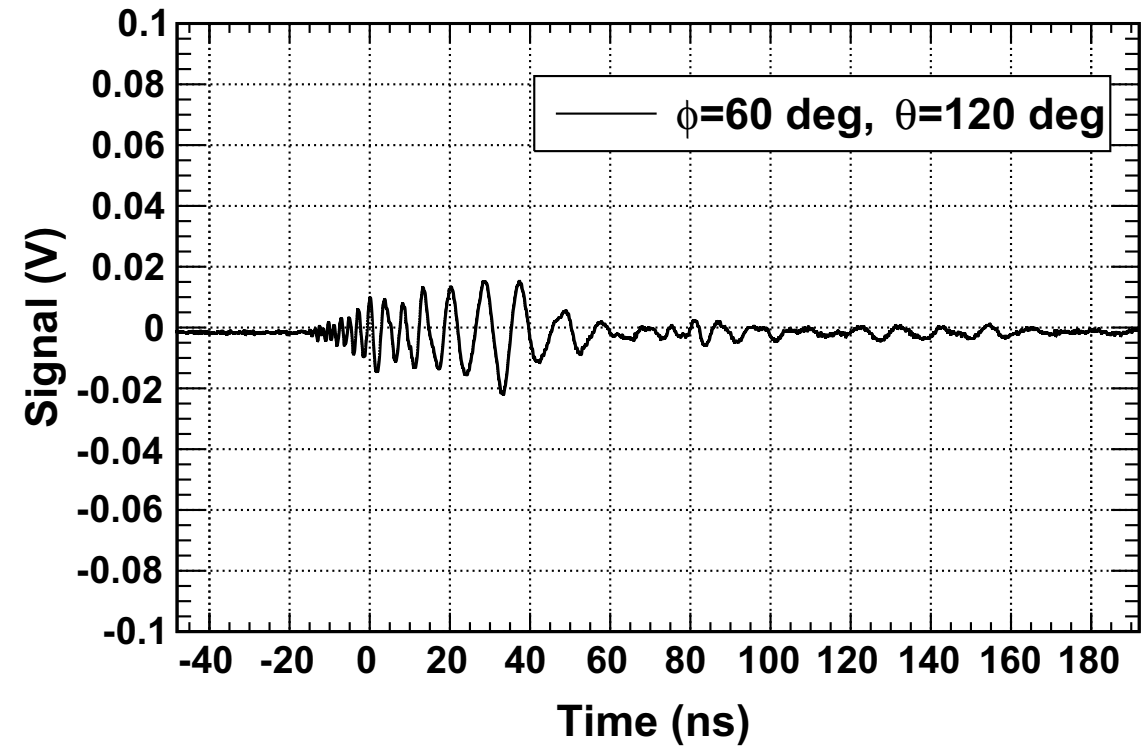

**Fig. 7.** Similar to Fig. 6, except at $\phi = 60°, \theta = 120°$.

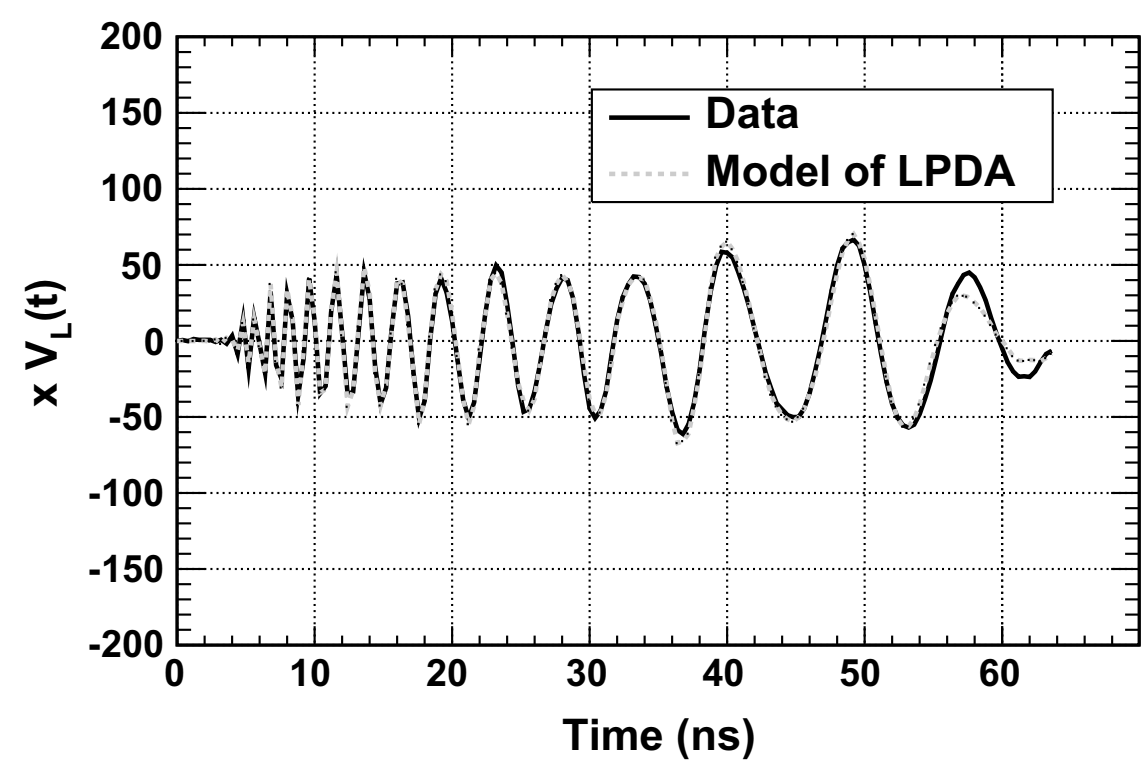

**Fig. 8.** The algorithm finds a solution for the effective height, $h(t)$, vs. time. The solution found by the algorithm (model of LPDA), solves the equation $xV_L(t) - h \circ h \circ \dot{V}_{src}(t) = 0$, in order to match the anechoic chamber data (data). The algorithm terminates after the mean squared-difference between model and data (the score) has improved by a factor of $\approx 100$.

## 5. Confirmations of measurements

Although the solutions for the LPDA response converge such that they match laboratory data, it is necessary to show that response solutions explain multiple independent measurements. Section 5.1 demonstrates that anechoic chamber measurements with the ARIANNA low-noise amplifier can be explained with the impulse response of the amplifier, combined with the that of the LPDA. Section 5.2 involves independent measurements of the temperature-averaged RF attenuation length of the ice beneath the deployed ARIANNA stations in Moore's Bay.

In this section, pulses reflected from the ocean/ice interface are modeled using the ice properties and antenna properties. Time series data from the 2006–7 and 2013–14 Antarctic seasons are used. The 2013 oscilloscope data was observed to agree with the ARIANNA data acquisition system, a result that has been demonstrated before [25]. The measurements of the ice properties were made in the interim (2010–10 and 2011–12), and are calibrated to be independent of the antenna model. Section 5.3 uses data taken in 2010 in Aldrich Park at UC Irvine, with different antenna orientations.

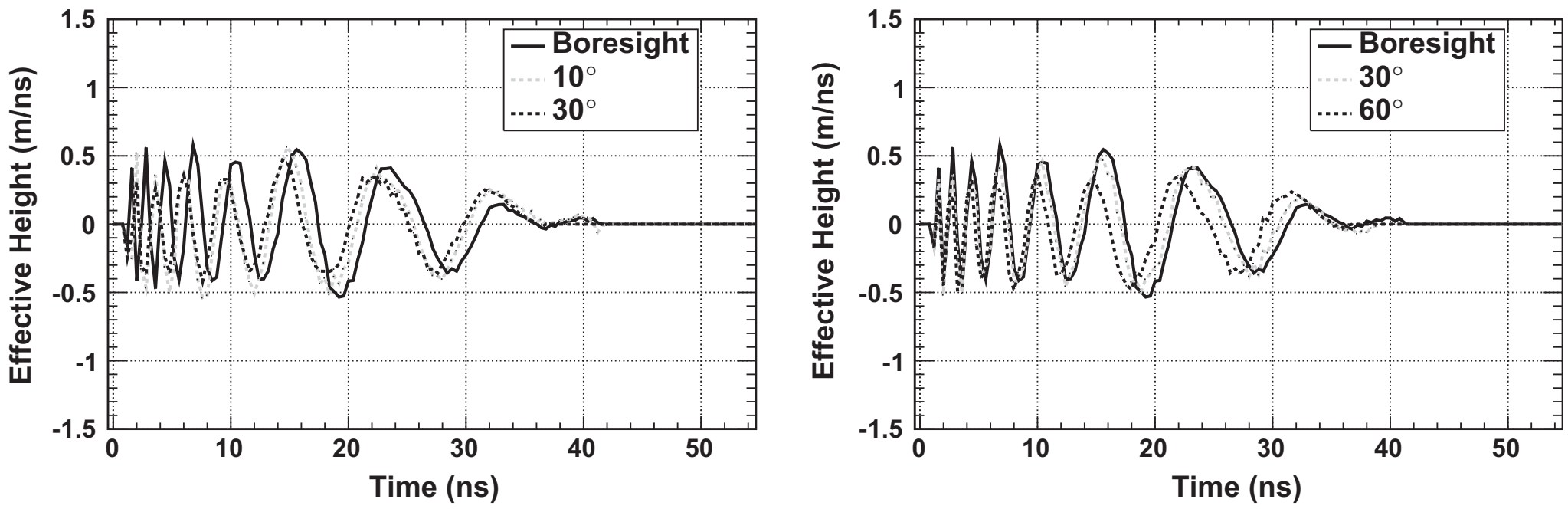

**Fig. 9.** The results for the effective height. (Left): selection of E-plane results, with $\phi$ varying, $\theta = 90°$. (Right): the same for the H-plane, with $\theta$ varying, $\phi = 0°$. In all cases the radiated wave polarization is parallel to the transmitter E-plane, which was kept fixed.

### 5.1. ARIANNA custom amplifier

The ARIANNA amplifier was designed to match the antenna signals to the ARIANNA waveform digitizers, while maintaining a high signal to noise ratio (a small noise figure). The specific design used here was developed in 2012, powered by 3.3 V DC, with a [50–1000] MHz bandwidth. The gain decreases with frequency (Fig. 10), so the impulse response function is not a $\delta$-function. To measure the impulse response, a 0.5-ns wide pulse was attenuated and fed directly into the amplifier. The output waveform was recorded, and from that waveform the impulse response was computed such that convolution of the response and the impulse produced the output waveform. The gain versus frequency, and impulse response are plotted in Fig. 10.

The amplifier was inserted into the anechoic chamber system, between the 1 GHz oscilloscope and the LPDA receiver, with a total of 36 dB of attenuation added in the system. The resulting data in Fig. 11 has been corrected for attenuators. The data matches the model, in amplitude and shape. The (Pearson) correlation coefficient, $\rho$, is used to quantify agreement between signals $x$ and $y$:

$$\rho = \frac{\mathrm{Cov}(x, y)}{\sqrt{\mathrm{Var}(x)\mathrm{Var}(y)}} \tag{13}$$

For the result in Fig. 11, $\rho = 0.89$. The equation for the modeled waveform is identical to that of the anechoic chamber system (Eq. (9)), with the amplifier impulse response (denoted $A$) added. (The constant numerical factors in front of Eq. (14) commute with the convolution operators). If either the amplifier transfer function, or the LPDA response function are not included, then the correlation coefficient drops to insignificant levels. The model amplitudes for early times, corresponding to higher frequency content, are slightly smaller than the data. This has little effect on $\rho$, however, because $\rho$ values are driven by the most powerful amplitudes.

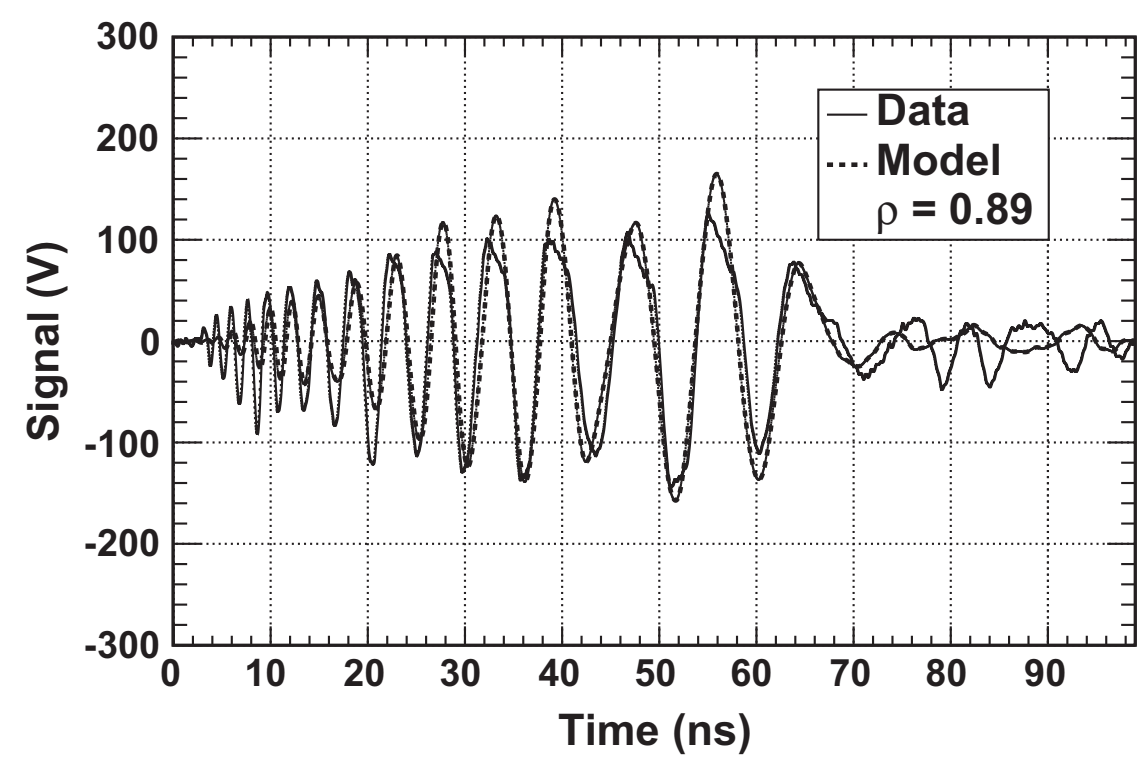

**Fig. 11.** Anechoic chamber amplified data, with model. The recorded data is shown as a solid line, and the dashed line is the model from Eq. (14). Agreement between the model and data is considered good because $\rho = 0.89$.

$$V_L(t) = \frac{1}{2\pi rc}\left(\frac{Z_0}{Z_L}\right) A \circ h_{rx} \circ h_{rx} \circ \dot{V}_{src}(t) \tag{14}$$

### 5.2. Ice shelf data

The RF attenuation length of the ice shelf in Moore's Bay has been measured several times (c.f. [5,20]), and will be updated in a forthcoming publication. In each radio echo measurement, a calibration pulse is recorded by the receiver and transmitter system. The antennas are then pointed down, and the recording oscilloscope is triggered on a delay such that it records the reflection from the ice/ocean interface beneath. By comparing the calibration pulse to the reflected pulse, the attenuation length is measured. The shelf depth is extracted from the oscilloscope time delay that captures the reflection. The shelf-depth obtained this way is $576 \pm 8$ m. The attenuation length $\lambda(\nu)$ causes the electric field mode $E(\nu)$ at frequency $\nu$ to be absorbed in the ice as $\propto \exp(-r/\lambda(\nu))$. The introduction of the reflection coefficient for electric fields accounts for the ocean/ice interface. Table 1 summarizes linear fits to the attenuation length vs. frequency data over several seasons in Moore's Bay, assuming 100% reflection at the ocean surface.

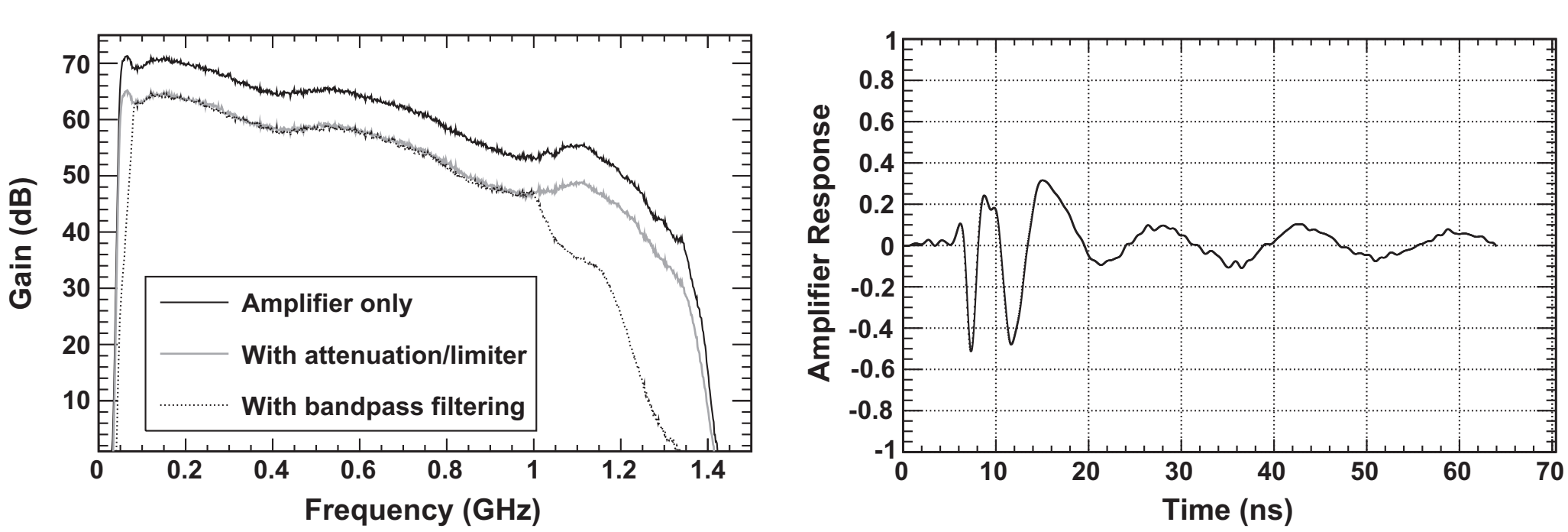

**Fig. 10.** (Left): The gain vs. frequency of the ARIANNA amplifier. (Right): The impulse response of the ARIANNA amplifier. The $y$-axis has been normalized such that the autocorrelation of this response is 1.0.

The linear fits to the data from all seasons are consistent, as demonstrated in Table 1. The attenuation lengths are conservative, because some of the returned power loss could be attributed to a non-ideal reflection coefficient. Prior measurements indicate that the interface in Moore's Bay is smooth compared to the shelf ice near the coast, [7,26]. Newer measurements of the reflection coefficient $\sqrt{R}$ indicate that $\sqrt{R} \approx 1$ [5,6], with almost negligible surface roughness. A more detailed discussion of the ice properties of Moore's Bay can be found in [5], and in forthcoming publications. Following convention, $R$ is the reflection coefficient for signal power, and $\sqrt{R}$ for electric field amplitudes. To model the reflections from the ocean, Eq. (14) is applied, with an additional step to account for the frequency-dependent attenuation length, and a reflection coefficient (Eqs. (15)–(17)). The 180-degree phase-shift caused by reflection between ice and salt water requires an overall minus sign.

$$E(t) = \left(\frac{1}{2\pi rc}\right)\left(\frac{Z_0}{Z_L}\right) h_{rx} \circ \dot{V}_{src}(t) \quad (15)$$

$$\widetilde{E}(\nu) = F_\nu(E(t)) e^{-\lambda(\nu)/r} \quad (16)$$

$$V_L(t) = -\sqrt{R} h_{rx} \circ F_\nu^{-1}(\widetilde{E}(\nu)) \quad (17)$$

Eq. (15) is just Eq. (7) for the matched-impedance case, and the input pulse $V_{src}$ was typically a 2.5 kV, 1 ns wide pulse from a Pockel Cell driver. In the 2006 measurement, the receiver and transmitter antennas were Seavey radio horns, and the measured effective height, at boresight, is shown in [10]. In the 2013 measurement, the receiving antenna was the ARIANNA LPDA. Because the attenuation length depends on frequency, the data is transformed to the Fourier domain, multiplied by the frequency-dependent attenuation factor, and transformed back to the time domain (Eqs. (16) and (17)). The operators $F_\nu$ and $F_\nu^{-1}$ represent the Fourier transform and inverse Fourier transform, respectively. Eq. (17) is just Eq. (3), in the matched-impedance case. No correction is made to $h_{rx}(t)$ for the effect of the surface snow, which has an index of $1.29 \pm 0.02$ (see below).

Fig. 12, left, presents the comparison of the 2006 data, and Fig. 12, right, the more recent data. A prototype of the ARIANNA analogue transient waveform digitizer (ATWD) was brought to Moore's Bay in 2010, and recorded similar results [25]. The linear fit $\lambda(\nu) = -(140 \pm 20)$ m/GHz $+(470 \pm 20)$ m was used for the attenuation length, and $\sqrt{R} = 1.0$, from Table 1 (These numbers are consistent with prior measurements, and will be updated in a forthcoming publication). Notice that, in either comparison, the negative sign in front of Eq. (15) is necessary to obtain the correct phase, indicating that the reflection at the ocean/ice interface is taking place between two materials with differing indices of refraction ($n_2 > n_1$).

**Table 1**
Attenuation length fit parameters, assuming 100% reflection at the shelf-bottom. The year and transmitter/receiver models (TX/RX) are shown in the first and second columns, respectively. The constant and slope of the linear fits are in the third and forth columns, respectively, and the $\chi^2/dof$ is shown in the fifth column. The form of the linear fit is $\lambda(\nu) = L_0 + \alpha\nu$. The 2006 data was published with 25 MHz bins, and has a larger $\chi^2/dof$. With 150 MHz bin-width, the $\chi^2/dof$ is reduced to 0.4. The final row indicates a fit to the average of all data collected.

| Year | TX/RX | $L_0$ (m) | $\alpha$ (m/GHz) | $\chi^2/dof$ |
|---|---|---|---|---|
| 2006 | S/S | 450 ± 10 | −120 ± 10 | 2.4 |
| 2011 | L/L, S/L | 475 ± 30 | −150 ± 60 | 0.42 |
| Ave. | L/L, S/L, S/S | 470 ± 20 | −140 ± 20 | 0.3 |

The correlation coefficients are larger than the naive expectation for signal correlation with noise: $\rho \approx 0$. Further studies of the thermal noise environment at the temperatures of Moore's Bay with the ARIANNA system indicate that the correlation between signal and realistic noise is $\approx 0.2$ for un-altered bandwidth, and $\approx 0.3$ for bandwidth-limited noise in the presence of band-pass filtering [27]. A good discussion of this subject can be found in [28]. Fig. 13 demonstrates that if the pulse from Fig. 12 is sent through the air, with the same equipment, the comparison still indicates agreement. That is, the quality of the model does not depend on the frequency dependent attenuation profile of the ice.

It is important to note that the snow plays no role in the agreement between model and data in Fig. 12. Several factors that could be corrected, due to the snow, are the medium conductivity, antenna impedance and the group delay. The antenna simulation package NEC4 [21] was used to reproduce in-air properties of the LPDA (such as impedance, VSWR, and radiation pattern), by solving for the antenna current using a method-of-moments approach. The simulation package allows the antenna to be embedded in a dielectric medium, with tunable index, $n$, and conductivity. The conductivity of Antarctic surface snow is small ($\approx 10$ μS/m) [29]. Varying the conductivity by an order of magnitude technically changes the antenna impedance, but not enough to effect $h_{rx}(t)$.

The simulated NEC antenna impedance is affected mainly by $n$: the real and imaginary parts of $Z_{in}$ are shifted down by a factor of $n$, and to lower frequencies: $\nu \to \nu/n$. However, the LPDA is designed for broadband, frequency-independent use, and the shift in lower cutoff frequency from 105 to 80 MHz has little effect on the structure of $h_{rx}(t)$ (derived from simulated parameters). The reduction by a factor $n$ in $Z_{in}$ could have caused a $\approx 15\%$ reduction in the amplitude of $h_{rx}(t)$ (which goes as $|\sqrt{Z_{in}}|$). However, the average $|Z_{in}|$ is 80Ω in the NEC4 model (agreeing with network analyzer measurements in air), whereas Eq. (9) assumes 50Ω. The measured $h_{rx}(t) \circ h_{rx}(t)$ must therefore be a factor of $(Z_L + Z'_{in})^2/(Z_L + Z_{in})^2$ larger, to compensate. With $Z_L = 50\Omega, Z'_{in} = 80\Omega$, and $Z_{in} = 50\Omega$, this factor is $1.3^2$, which is equal to the measured $n^2$. When the antennas are placed in snow (and each $h_{rx}(t)$ drops by $\sqrt{n}$), $Z_0 \to Z_0/n$. Thus, these effects in Eq. (9) *cancel* when the LPDA is placed in the snow. Ultimately, the empirical data matches the model, and it is safe to assume the role of the snow is negligible.

The NEC model was also used to check that the position of the LPDA does not strongly affect the results. The depth of the LDPA does not affect the impedance as long as $n$ is constant near the surface. Near the air/snow interface, this is most likely a first-order approximation. However, measurements of the VSWR in [19] reveal that the LPDA bandwidth extends down to 80 MHz, verifying the NEC model prediction near the surface. Because the front-to-back (F/B) ratio of the LPDA is −15 dB, surface effects likely do not matter as long as the antenna is facing downward. Density measurements of the surface firn vs. depth [5,20] indicate that treating the index as a constant for the first few meters is reasonable. Finally, because the group delay is the derivative of the S21 phase, it is a relative quantity should not be affected as long as the index is constant. As a final check, the simulated impedance, gain, and the theoretical group delay equation were folded into the standard formula for $h_{eff}(\nu)$ [30] and inverse Fourier-transformed to obtain $h_{rx}(t)$. The snow and air versions of $h_{rx}(t)$ had a correlation coefficient of 0.98 in this case.

### 5.3. Further comparisons in air

The time-domain model derived from anechoic chamber data must be checked for a variety of situations, including locations with no ice or snow, and different band-pass filtering. In 2010, data was taken in Aldrich Park to investigate the LPDA waveform shape for a variety of angles. Filters rejecting CW noise were applied.

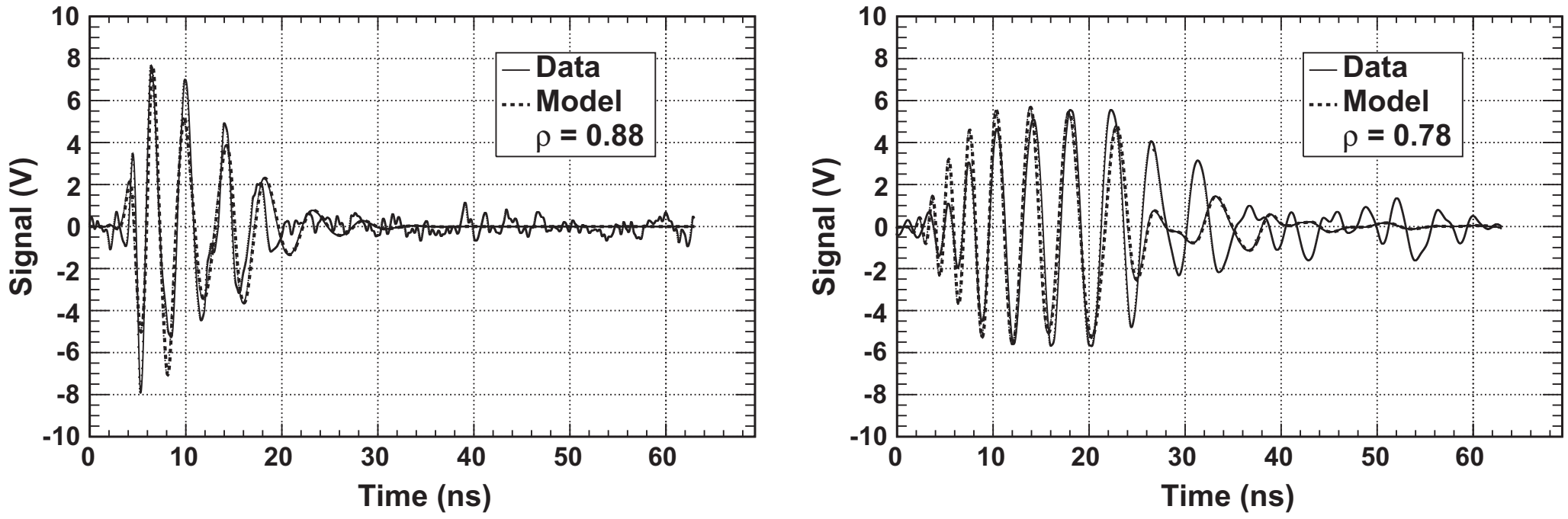

**Fig. 12.** (Left): 2006 Moore's Bay comparison. (Right): 2013 Moore's Bay comparison. The recorded data are shown as solid lines, and the model from Eq. (15) is shown as a dashed line. The value of $\rho$ is 0.88 for the 2006 data and 0.78 for the 2013 data, indicating agreement. The waveforms have been corrected for attenuators used to keep amplitudes in the linear range of amplifiers and oscilloscopes.

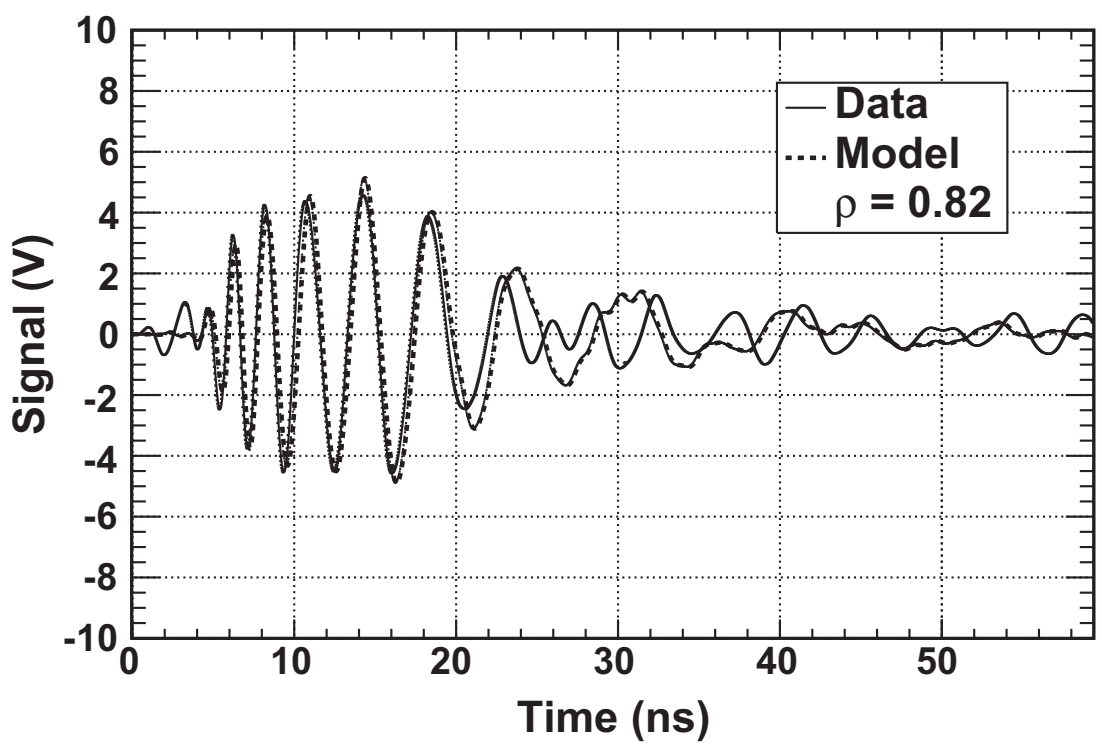

**Fig. 13.** In-air Seavey to LPDA calibration through 23 m of air. The recorded data is shown as a solid line, and the model from Eq. (15) is shown as a dashed line ($r = 23$ m, $\lambda = \infty$). The value of $\rho = 0.82$. The waveforms have been corrected for attenuators used to keep amplitudes in the linear range of amplifiers and oscilloscopes.

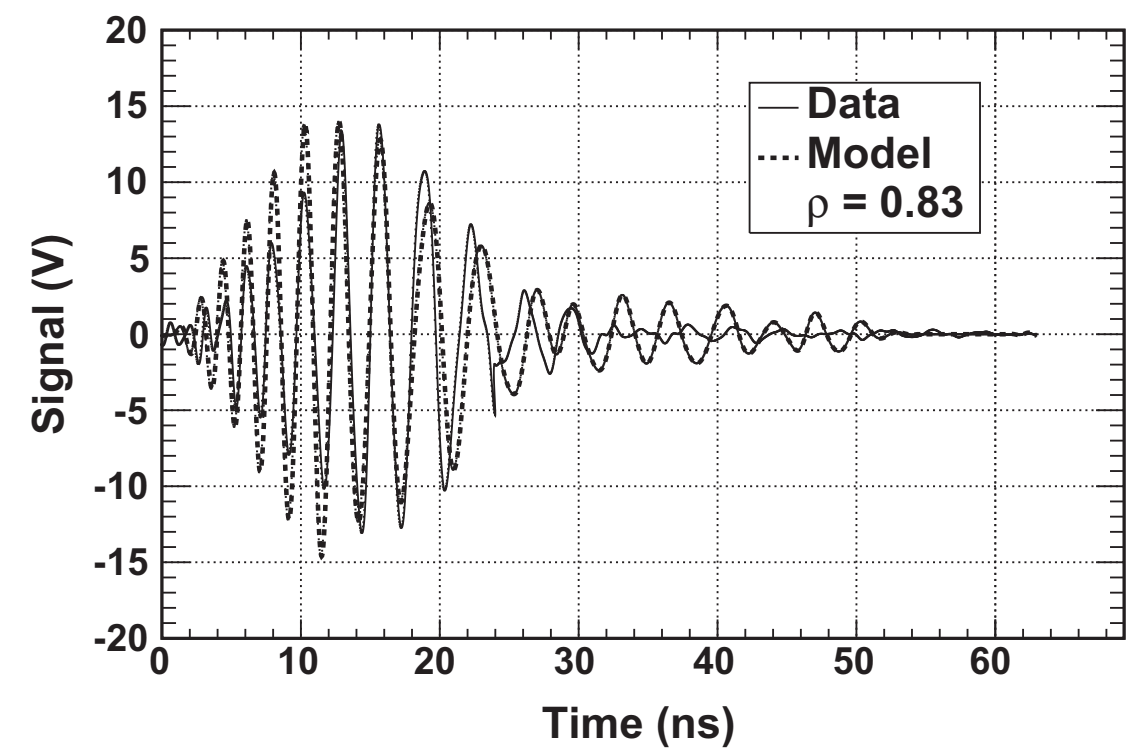

**Fig. 14.** A model of a pulse obtained in Aldrich Park in 2010, using two LPDA antennas, a PCD, and filters, with a separation of 5 m. The correlation coefficient is $\rho = 0.79$.

Specifically, Mini-circuits NHP-300+ and NLP-600+ filters were used to eliminate local radio stations (near 100 MHz) and emergency VHF bands, respectively. A Pockels Cell Driver (PCD) with a negative 1.2 kV, 1-ns pulse was transmitted to an LPDA, and received by an identical LPDA 5 m away. This PCD was different from the unit used in Moore's Bay.

Fig. 14 shows a comparison using the LPDA boresight effective height that accounts for filters, the 5 m separation, and attenuation (60 dB, required because of the large PCD amplitude). The correlation coefficient is $\rho = 0.83$, after removing a reflection that occurs 40 ns after the original pulse arrives in the oscilloscope. The comparisons in Sections 5.3, 5.2, and 5.1 are summarized in Table 2. For a variety of settings, the model reveals high correlation with measurements.

If the models represented in Figs. 11–14 were perfect, and the signals being modeled were free of noise, the correlation coefficients ($\rho$) in Table 2 would be 1.0. The noise in the data of Figs. 11–14 is bandwidth limited, and non-uniform in power per unit frequency. This type of noise lowers $\rho$ values to the levels of Table 2. Thus, systematic errors in the response model are not required to explain the correlation coefficients. For example, noise samples from the amplified anechoic chamber data prior $t = 0$ in Fig. 11 are used as a noise sample in that test. The correlation coefficient of the model output with itself, plus this noise component, is 0.85. Thus, the noise alone explains the $\rho$ value between model and data ($\rho = 0.89$). This procedure was repeated for each test (Figs. 11–14), and similar results to Table 2 were obtained. In each case, the noise sample was well separated in time from the signal, and the noise level produced the same signal to noise ratio as the data when added to the model.

**Table 2**
Correlation coefficients, $\rho$, determined from the different measurements described in the text.

| Exp. Setting | Fig. | $\rho$ |
|---|---|---|
| Chamber+amplifier | 11 | 0.89 |
| Ice sounding (Moore's Bay 2006) | 12 | 0.88 |
| Ice sounding (Moore's Bay 2013) | 12 | 0.78 |
| In-air over ice (Moore's Bay 2012) | 13 | 0.82 |
| In-air (Aldrich Park 2010) | 14 | 0.83 |

## 6. Askaryan pulses in ARIANNA data

Given an understanding of how the Askaryan electric field would be transformed in the data acquisition, the next logical step is to predict properties of the signal based on a theoretical understanding of the neutrino interaction. Various authors have studied the problem theoretically [13,14,31], and several experimental confirmations have been achieved [32,33]. The experimental observations recreate the ultra-high energy electroweak interaction by building a cascade energy equivalent to the expectation for a cosmogenic neutrino. The total energy is equal to the sum of the separate energies of the charged particles in a beam built from photons or electrons.

The negative charge excess that develops over several meters in the dielectric material leads to RF radiation, and the shape of the pulse reflects the details of the longitudinal development of charge, and perspective from which the pulse is observed, since the highest frequencies are confined to the Cerenkov angle. In Section 6.1, signal predictions are shown with respect to the observation angle relative to the Cerenkov angle. In Section 6.2, the result of varying the angle at which the Askaryan pulse interacts with the ARIANNA antenna is presented.

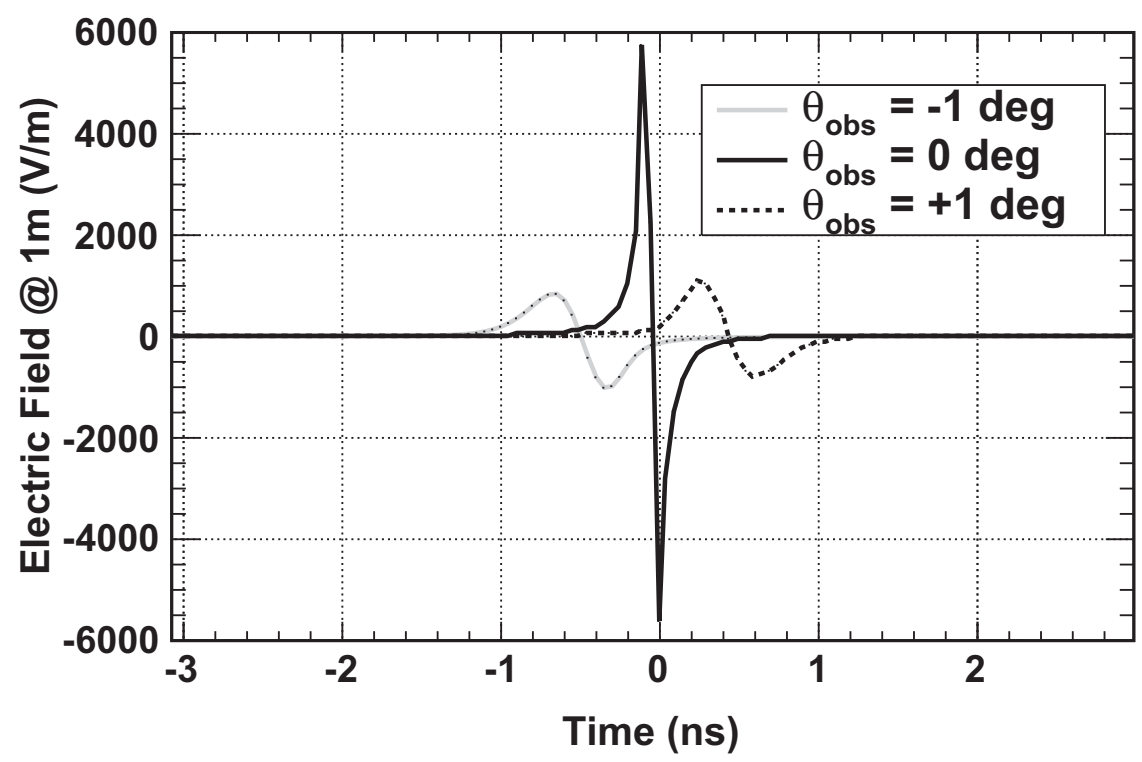

**Fig. 15.** The electric field pulse created by a 100 TeV event, scaled to an energy of 10 EeV, $\theta_{obs} = -1^\circ, 0^\circ$ and $1^\circ$.

### *6.1. Prediction vs. observation angle*

A description of how Askaryan electric fields are generated from the negative charge excess profile versus cascade depth can be found in [13,14], and early Monte Carlo studies in [34]. Several key observations arise from these calculations. First, the electric field amplitude (and vector potential) scales with the total energy, due to coherence effects. Second, the shape of the electric field corresponds to a temporal derivative with respect to the retarded time of the excess negative charge profile. Because the charge excess is moving in the lab frame, the pulse shape traces the derivative of the excess charge with respect to cascade depth. The LPM effect [35] stretches the charge excess profile above $10^{16}$ eV, and therefore stretches the pulse, while reducing the amplitude. Secondary peaks in the charge distribution and therefore the electric field are also caused by the LPM effect. Because the pattern of energy deposition varies from event to event in the strong-LPM regime, there can be considerable event-to-event variation in the radio frequency spectrum.

Fig. 15 displays Askaryan pulses derived from the ZHS Monte Carlo [34], with subsequent modifications for non-cylindrical form factors of the charge distribution [14]. The electric fields are scaled to correspond to a total energy of 10 EeV, but were created from charge excess profiles from 100 TeV events (the scaling is linear). Because the LPM effect becomes relevant above $10^{16}$ eV [12], the pulses are unaffected by it, and are smooth and unstretched. The center of graph corresponds to a retarded time of zero. The pulse asymmetry (difference in maximum and minimum values) for non-zero observation angles (color scale) is caused primarily by the non-symmetric charge distribution. For negative observation angles, the retarded time dictates that the end of the charge excess profile is observed *first*, and the electric field is anti-symmetric for reflections across the y-axis where the retarded time is zero.

The time-domain response of both the ARIANNA LPDA and the low-noise amplifier produce results that match data. Theoretical Askaryan pulses can be combined with them to produce experimental predictions for the neutrino signal, or *templates*. For electric fields not subject to the LPM effect, the accuracy of the templates is limited only by model and experimental uncertainties. Electric fields subject to the LPM effect have also been studied, to establish how the templates change. However, this study is not meant to be comprehensive. Assuming a matched coaxial cable, and including amplifier effects, Eq. (3) reduces to $V_L(t) = A(t) \circ h_{rx}(t) \circ E_\nu(t)$, where $A(t)$ again represents the transfer function of the amplifier. Fig. 16 shows the signal templates $V_L(t)$, while varying the observation angle. Technically, ice absorption is also taken into account over a 1000 m path length, however, it is shown below that this has a negligible effect on waveform structure.

While the signal amplitudes are predicted by the analysis, the waveforms have been scaled such that the maximum voltage is 1.0, so that shapes can be compared (Fig. 16). The only change in the templates with respect to observation angle, besides overall amplitude, is the frequency content early in the wave. Higher frequency modes are expressed early in the wave as the observation angle decreases to zero, however these modes are lost off-cone. The structure of the LPDA response causes the high frequency modes to be recorded earlier, because the smaller dipole elements are located nearer to the antenna feed point. (The next section shows templates for non-boresight angles).

The LPM versions of the templates demonstrate a stronger dependence on the observation angle than the non-LPM versions. There are three main effects. First, the on-cone (0°) versions are identical, because the electric fields are identical. That is, an extreme narrowing of the Cerenkov cone width is irrelevant, if the observation is taking place at precisely the Cerenkov angle. Second, for off-cone observations, the high frequency modes near the beginning of the waves are suppressed, because those modes are suppressed in the original electric field. Finally, the lower frequency modes (near the end of the waves) appear to be enhanced relative to the 0° case. This is an artifact of the normalization, which sets the maximum voltage (whenever it occurs) to 1.0. In the un-normalized templates, oscillations between 20 and 30 ns from the 0° case are the largest in the set. It is important to mention here that the LPM-templates shown in Fig. 16 are not meant to be comprehensive, and demonstrate the need for further analysis.

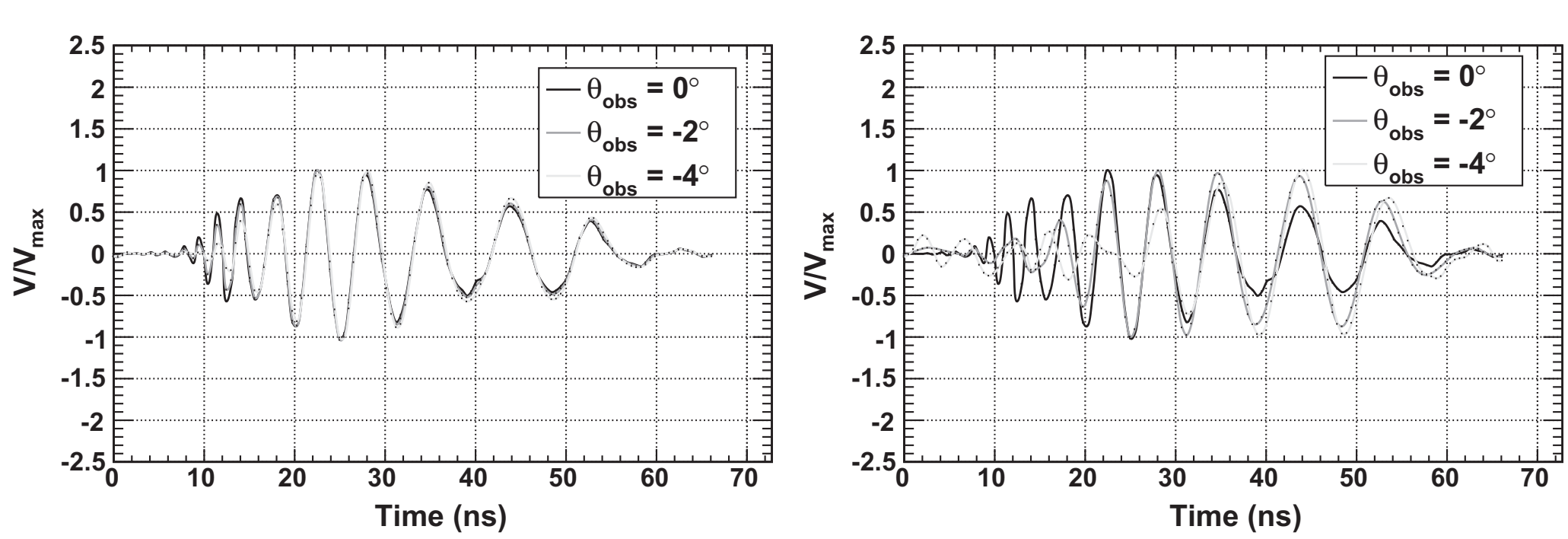

**Fig. 16.** Signal templates for a 10 EeV neutrino, via the Askaryan effect. The LPDA observes the cascade at boresight, and the results for observation angles of 0°, −2° and −4° are shown. (Left): Signals generated by NC neutrino interactions, appropriate for CC interactions in the absence of significant LPM suppression. (Right): the same events, with the LPM effect included, appropriate for electron-type CC interactions.

Correlating templates with each other produces an interesting result: if the E-plane angle (next section) is restricted to two beam-widths or less ($\approx 60^\circ$), the correlations between templates obey $\rho \geqslant 0.8$. This result includes varying the E- and H-plane angles, and the observation angle. Because 0.8 is much larger than auto-correlation coefficients produced by noise in ARIANNA analyses, typically $\rho \lesssim 0.3$), it is acceptable to use the time-dependent pulse generated from an electromagnetic cascade at energies below the influence of the LPM, and scale it to higher energies to predict the waveforms from the dominant hadronic cascades. LPM-dominated events (CC events with energies greater than 1 EeV) make up 20%–30% of the total event rate [12] in ARIANNA, and do not obey $\rho \geqslant 0.8$, for correlations between all templates from the LPDA forward lobe. To understand the final fraction of events which do undergo the LPM effect, future development of this work will extend template production to include LPM physics.

### 6.2. Prediction vs. antenna angle

Although the effect of the observation angle is interesting theoretically, the majority of detectable signals in the ARIANNA system will be electric fields with small observation angles, above the detector threshold. The effect of the LPDA on the signal must be clear, since this effect (and of the amplifier) must be deconvolved to reveal the electric field. Fig. 17 demonstrates the effect of the LPDA responses from Fig. 9, and the amplifier, on an on-cone Askaryan pulse at 10 EeV. In each case, the signal is assumed to be co-polarized. The cross-polarization fraction [5,6], which measures how much power leaks into the cross-polarized direction due to ice propagation, is $\leqslant 5\%$ for the data in ice soundings taken in Moore's Bay. Fig. 17 retains the properties of Fig. 9. The LPDA response approaches uniformity within the main lobe of the E- and H-planes, especially at lower frequencies.

Although ice absorption over the total path length affects the overall amplitude in a triggered event, it has been checked that the effect on the waveform shape is small (Fig. 18). The slope of the measured attenuation length vs. frequency is not steep enough to produce a difference comparable to the shelf depth over a few hundred MHz. The differences in high and low frequency attenuation do not have enough time to warp the waveform shape. With an average (measured) shelf depth of $576 \pm 8$ m, the attenuation effect works out to approximately a $-16$ dB/km overall scale factor, which is nearly independent of frequency.

### 6.3. Confirmation of ARIANNA Monte Carlo amplitudes

The ARIANNA Monte Carlo simulation, ShelfMC, predicts the overall exposure and sensitivity to neutrino flux, given a variety of parameters [12]. Natural factors, such as the chemical composition of cosmic rays and the distribution of cosmic ray sources with respect to redshift [36], must be determined independently or taken as free parameters. However, once the neutrino flux interacts in the ice, it produces a cascade. The cascade electric field strength versus frequency (in ice) has been measured experimentally [32] for cascades of equivalent total energy to GZK events. The ShelfMC simulation has been adapted from that of prior experiments [9,10]. These Monte Carlo simulations multiply the electric field strength (parameterized as in [37]) with an effective height equation in the following way:

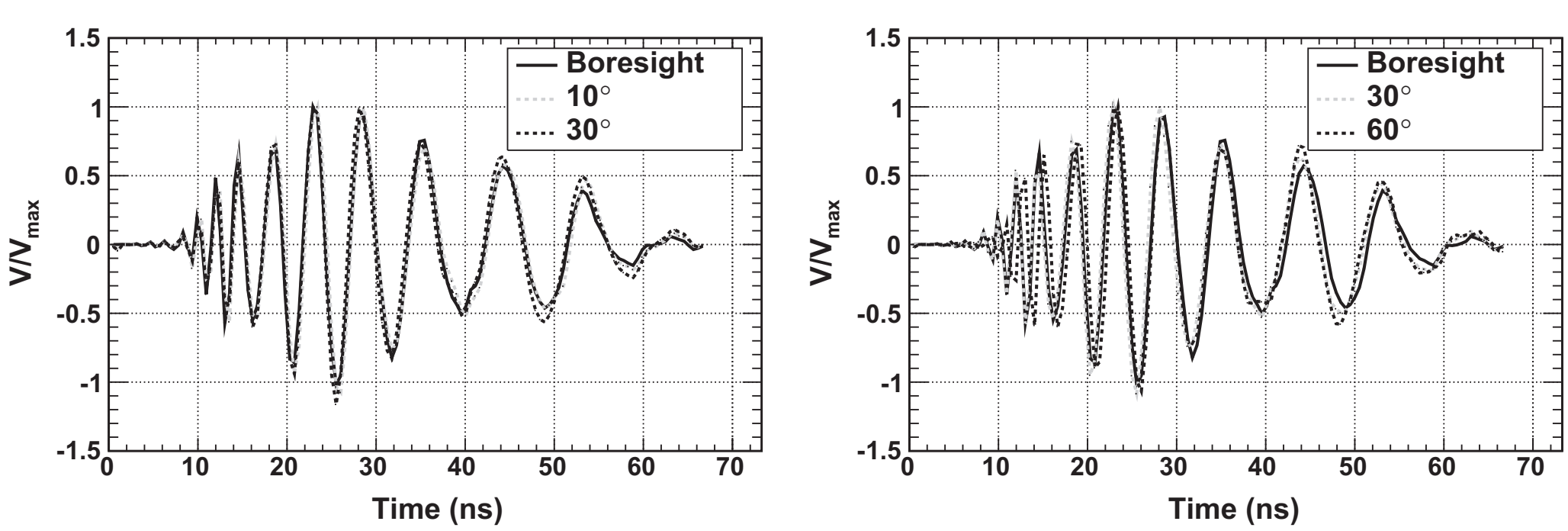

**Fig. 17.** Signal templates for a 10 EeV neutrino, via the Askaryan effect. The LPDA observes the cascade at varying angle with respect to boresight, and the observation angle is kept constant at $0^\circ$. (Left): Varying antenna angle in the E-plane only. (Right): Varying the antenna angle in the H-plane only.

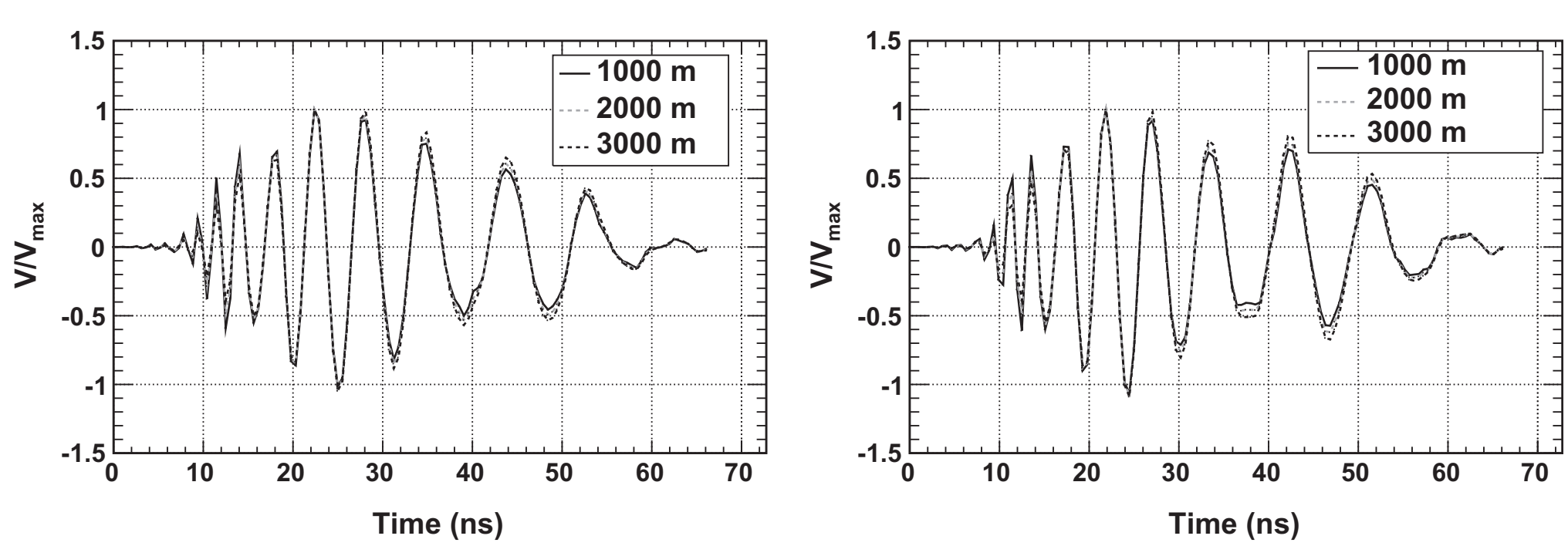

**Fig. 18.** These signal templates includes the usual suite of electronics effects (LPDA, filters, and amplifier), and the effect of the ice. The measured attenuation length vs. frequency is $\lambda(\nu) = (470 \pm 20) - (140 \pm 20)\nu[\text{GHz}]$ (meters). The path length is varied and shown in the legend, and the shape changes accordingly. The waveforms are normalized such that the maximum voltage is 1.0. (Left): LPDA boresight. (Right): $60^\circ$ in the H-plane.

$$E_\nu^{1m}\left[\frac{V}{\text{m MHZ}}\right] = \left(\frac{f}{f_0}\right)\frac{E[\text{TeV}]\times 2.53\times 10^{-7}}{1+(f/f_0)^{1.44}}\left(\frac{\sin\theta_{obs}}{\sin\theta_C}\right) \quad (18)$$

$$E_\nu(\theta_{obs},r,f) = E_\nu^{1m}(f)\exp\left(-\ln 2\left(\frac{\theta_{obs}-\theta_C}{\Delta\theta(f)}\right)^2\right)\frac{e^{-r/\lambda}}{r} \quad (19)$$

$$V_{ant} = \frac{\sqrt{R}\Delta f}{2\sqrt{2}}\sum_i h_{eff}(f_i)(f_{had}+f_{em})E_\nu(\theta_{obs},r,f_i)G(\theta_E,\theta_H) \quad (20)$$

In Eq. (18), the basic electric field strength is shown, and $f_0 = 1150$ MHz. In Eq. (19), $\theta_{obs} - \theta_C$ measures how far the observation is from the Cerenkov cone, $\Delta\theta$ is the width of the Cerenkov cone, and $\theta_C = 56^\circ$ is the Cerenkov angle in ice. The cone width $\Delta\theta$ depends on frequency, and whether the event is electromagnetic or hadronic in nature. Eq. (21) below gives the electromagnetic dependence, which depends on the LPM effect. In Eq. (19), $r$ and $\lambda$ are the distance from the neutrino vertex and ice attenuation length, respectively. In Eq. (20), $f_{em}$ and $f_{had}$ are the fractions of energy in the electromagnetic and hadronic component of the neutrino cascade, and the function $G$ averages (in-quadrature) Gaussian models of the forward lobe of the LPDA radiation pattern in the E- and H-planes. If the event reflects from the ocean, then $R$ is the reflection coefficient for power. Finally, $h_{eff}$ is the scalar expression for antenna effective height at a given wavelength [30].

$$\Delta\theta(E_\nu,f) = 2.7^\circ\frac{f_0}{f}\left(\frac{E_{LPM}}{0.14E_\nu + E_{LPM}}\right) \quad (21)$$

The convolution theorem states that multiplication of two functions in the Fourier domain is identical to convolving those functions in the time domain. Eq. (20) multiplies the scalar effective height formula for a single frequency ($h_{eff}$) with the electric field. Thus, Eq. (20) is like the proper convolution of the electric field and antenna response ($h_{rx}(t)$) with the complex phase factor neglected. For pure ice, $E_{LPM} \approx 0.3$ PeV [35], depending on the ice density, and the cascade begins to elongate near 2 PeV. The nominal ice density of 0.92 g/cc is assumed here. The frequency $f_0$ is the same as above. Antenna beam widths of 60° and 120° are used as the Gaussian widths in $G$ in Eq. (20), for the E- and H-planes respectively, and $\theta_E$ and $\theta_H$ are the incoming angles in the respective antenna planes. This description of the antenna main lobe is an approximation best suited for directions within one beam-width of the forward direction.

The full calculation of the antenna voltage yields a number meant to be compared to rms voltage fluctuations from thermal noise. The fractional deviation of ShelfMC from the maximum voltage in this work's model is shown in Fig. 19, with $\theta_E$ and $\theta_H$ varied through all the angles measured in the anechoic chamber, subject to $\theta_E < 60^\circ$. The convolution method from this work is denoted TD for time-domain. The Askaryan pulse comes from [14], using $E_\nu = 3\times 10^{18}$ eV, $\theta_{obs} - \theta_C = 0.3^\circ, f_{em} = 1$. No distance, reflection, or attenuation effects were applied to either the TD or ShelfMC numbers, since these effects equally shift both distributions. Thus, a comparison can be made between the ShelfMC expression for the maximum voltage, and the maximum voltage of the signal templates.

Fig. 19 demonstrates that the fractional difference between ShelfMC and TD is typically 10%. The $\chi^2/dof$ indicates a good fit. The excess near −0.4 comes from an overestimation by ShelfMC, where the Gaussian functions $G$ describing the radiation pattern over-predict the LPDA relative gain (Eq. (20)). In reality, the Gaussian approximation is valid only for angles well within one or two beam widths in the E-plane (about 60°, centered on forward direction), because the real radiation pattern (Fig. 4) decreases more quickly. Fig. 19 indicates that the errors in the signal to noise $V_{ant}/V_{thermal}$ from ShelfMC are modest (for constant $V_{thermal}$) when the more realistic TD model is employed. The TD model incorporates antenna phase effects, and even predicts the waveform shape (the templates). The thermal fluctuations are characterized by the *system temperature*, which remains constant because the antenna aperture and bandwidth are constant in either case.

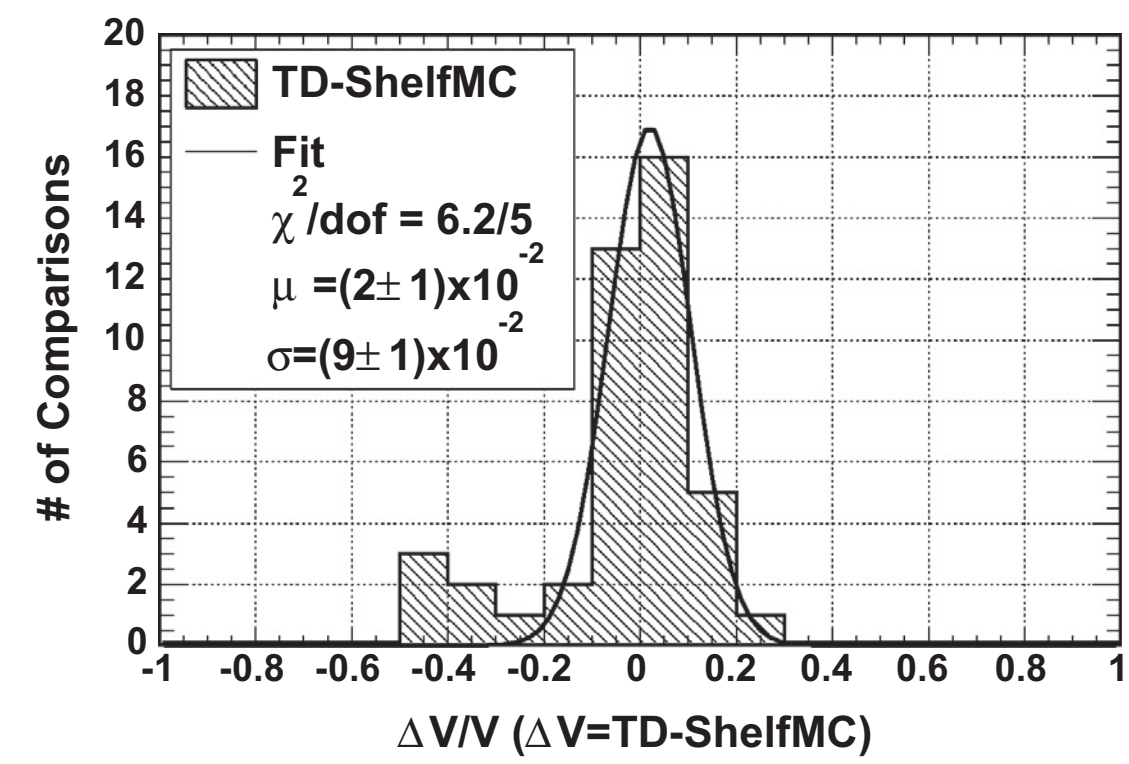

**Fig. 19.** The fractional voltage difference between ShelfMC (which relies on the formalism of Eqs. (18)–(20)) and the TD model of this work. The Askaryan signal is convolved with the each LPDA $h_{rx}(t)$ measured in the anechoic chamber, for each $\theta_E, \theta_H$ from the chamber turn-table positioning system. A restriction of $\theta_E < 60^\circ$ has been applied (approximately two beam widths), for a total of 43 comparisons.

## 7. Summary and conclusions

The RF response of the ARIANNA DAQ has been measured, with the purpose of predicting and quantifying the signatures from high energy neutrino interactions in Antarctic ice. This includes an iterative procedure to compute the effective height of the log-periodic dipole arrays (LPDA) serving as receiving antennas for the system, and the amplifier transfer function of the amplifier feeding the antenna signal to the digitizer. The LPDA response was determined in air, but prior work [19] and evidence presented in this paper demonstrates that the response remains valid for LPDA embedded in low density firn less than a few meters from the surface. This work has defined a procedure to compute the time-dependent signal Askaryan signal from the RF response, and this calculation is used to produce signal templates in several examples. The measured response is confirmed by data from multiple field contexts. The level of agreement between predicted and observed time domain waveforms produced correlation coefficients larger than 0.79. The predicted time-dependent wave packet is approximately 40 ns wide, changing shape according to the observation angle with respect to the Cerenkov cone, and the arrival direction with respect to the antenna. The change in shape due to the arrival direction is smooth and continuous in the forward lobe of the LPDA.

The example neutrino templates were computed from theoretical calculations of the time dependent signals generated from electromagnetic cascades [13,14,37] and convolved with the RF response function of the detector. The correlation coefficient between templates obeys $\rho \geqslant 0.8$, for variation of the E- and H-plane angles, and the observation angle. This result indicates that imprecise knowledge of the viewing angle has little impact on the overall form of the waveform, because the duration of electromagnetic pulse is short compared to the time scales of the system response. For similar reasons, there is little variation in the time-dependent waveform from hadronic or electromagnetic cascades, except in the case where the LPM effect is strong. Thus, it is acceptable to use the time-dependent pulse generated from an electromagnetic cascade at energies below the influence of the LPM, and scale it to higher energies to predict the waveforms from the dominant hadronic cascades. Though LPM-dominated events are sub-dominant in ARIANNA, [12], future development of this work will extend the template production to include a broader range of LPM associated physics. Finally, it was also shown that the additional frequency dependence introduced by ice attenuation has negligible impact on the time dependent shape of the waveform for interaction distances relevant to ARIANNA.

The ARIANNA digitizers operate via a positive and negative voltage threshold trigger that separates random thermal noise from signal-like templates like those presented in this work. Rates are kept low, while maintaining low voltage thresholds, by requiring that a signal rise above a positive threshold and drop below a negative threshold. Such a system is triggered by a generic bi-polar waveform. While there are risks to searching for a specific signal shape in data, the potential benefits include highly efficient thermal-noise trigger rejection. It has been shown in this work that the signal templates are bi-polar, and contain high-frequency (1 GHz) content in spite of ice absorption and antenna effects. Further, an understanding of the energy-scaling of the signal is required for energy measurement. The template analysis provides this benefit, confirmed at the 10%-level by Monte Carlo simulation (for events near the Cerenkov angle).

Finally, correlations between all computed templates (for varying Cerenkov observation angles, and incoming angles in the LDPA main lobe) remain above 0.85, implying that the templates are consistent enough to use in ARIANNA data analysis routines [27], without *a priori* knowledge of the observation or incoming angles. By comparing the average $\rho$-value between data channels and templates, random thermal triggers are rejected for entire seasons of data with relatively few complex calculations. Future directions along the lines published here include studying cross-polarized measurements to constrain Askaryan field polarization, and the production of templates corresponding to high-energy cosmic rays. The latter effort involves use of the CoREAS code [38,39].

## Acknowledgements

We wish to thank the staff of Antarctic Support Contractors, Raytheon Polar Services and Lockheed Martin, and the entire crew at McMurdo Station for excellent logistical support. We would like to acknowledge and thank the CReSIS project and the Anechoic Chamber facility management for the use of the world class anechoic chamber at the University of Kansas. This work was supported by generous funding from the Office of Polar Programs and Physics Division of the US National Science Foundation, grant awards ANT-08339133, NSF-0970175, and NSF-1126672.

## Appendix A. Further equations

This section reviews the definition of the antenna effective height used in Section 2, and the assumption that $Z_{in} \approx Z_L$. Eq. (9), which produces the result for the signal recorded in the anechoic chamber (used to solve for the antenna effective height), has been simplified assuming $Z_{in} \approx Z_L$. The signal $V_L(t)$ recorded in a line impedance-matched configuration on an oscilloscope, in response to an incident electric field $E(t)$, is given by Eq. (3):

$$V_L(t) = 2\left(\frac{Z_L}{Z_L + Z_{in}}\right) h_{rx}(t) \circ E(t) \tag{A.1}$$

The operator ($\circ$) refers to convolution. The left-hand side of this equation is strictly real. Taking the imaginary part of both sides:

$$\mathrm{Im} V_L(t) = 2\mathrm{Im}\left(\frac{Z_L}{Z_L + Z_{in}}\right) h_{rx} \circ E(t) \tag{A.2}$$

$$0 = \mathrm{Im}\left(\frac{Z_L}{Z_L + Z_{in}}\right) \tag{A.3}$$

$$0 = \mathrm{Im}\{Z_L Z_{in}^*\} \tag{A.4}$$

$$\mathrm{Im} Z_L \mathrm{Re} Z_{in} = \mathrm{Im} Z_{in} \mathrm{Re} Z_L \tag{A.5}$$

From the last statement, it follows that the phases of $Z_L$ and $Z_{in}$ must be equal. If the real parts are equal, then the imaginary parts are equal as well, and $Z_{in} = Z_L$. For most RF equipment, the real part is just 50Ω, and this is true for the cables in this work. As discussed in Section 5.2, however, the real part of the LDPA impedance is 80Ω on average. When placed in snow with an index of refraction $n = 1.3$, however, $|Z_{in}|$ drops to $\approx 50\Omega$, and the assumptions hold.

For an impulsive electric field, described by $E(t) = E_0\delta(t - t_0)$, the voltage read out by the antenna is

$$V_L(t) = 2\left(\frac{Z_L}{Z_L + Z_{in}}\right) h_{rx}(t) \circ E(t) = h_{rx} \circ E_0\delta(t - t_0) \tag{A.6}$$

$$V_L(t) = E_0 h_{rx}(t) \circ \delta(t - t_0) = E_0 h(t - t_0) \tag{A.7}$$

The antenna cannot reproduce the impulsive signal, and instead reads out a copy of the response function $h_{rx}$ at a time $t - t_0$, with the units of volts versus time, proportional to the electric field amplitude.

## References

[1] K. Greisen, Phys. Rev. D 16 (1966) 748–750, http://dx.doi.org/10.1103/PhysRevLett.16.748.
[2] G.T. Zatsepin, V.A. Kuz'min, JETP Lett. 4 (1966) 78–80.
[3] V.S. Berezinsky, G.T. Zatsepin, Phys. Lett. B 28 (1969) 423.
[4] S.W. Barwick, J. Phys. Conf. Ser. 60 (2007) 276–283, http://dx.doi.org/10.1088/1742-6596/60/1/060.
[5] J.C. Hanson, The performance and initial results of the ARIANNA prototype (Ph.D. thesis), University of California at Irvine, 2013.
[6] J.C. Hanson et al. (The ARIANNA Collaboration), Proceedings of the 32nd International Cosmic Ray Conference, Beijing, China. doi:http://dx.doi.org/10.7529/ICRC2011/V04/0340.
[7] C.S. Neal, J. Glaciol. 24 (1979) 295–307.
[8] G. Askaryan, Sov. Phys. JETP 14 (1962) 441–443.
[9] I. Kravchenko et al. (The RICE Collaboration), Phys. Rev. D 85. doi:http://dx.doi.org/10.1103/PhysRevD.85.062004.
[10] P.W. Gorham et al., The ANITA Collaboration, J. Astropart. Phys. 32 (2009) 10–41, http://dx.doi.org/10.1016/j.astropartphys.2009.05.003.
[11] S. Klein, Nucl. Phys. Proc. Supl. 229–232 (2012) 284–288, http://dx.doi.org/10.1016/j.nuclphysbps.2012.09.045.
[12] K. Dookayka, Characterizing the search for ultra-high energy neutrinos with the ARIANNA detector (Ph.D. thesis), University of California at Irvine, 2011.
[13] J. Alvarez-Muniz, A. Romero-Wolf, E. Zas, Phys. Rev. D doi:http://dx.doi.org/10.1103/PhysRevD.81.123009.
[14] J. Alvarez-Muniz, A. Romero-Wolf, E. Zas, Phys. Rev. D 84 (2011) 103003, http://dx.doi.org/10.1103/PhysRevD.84.103003.
[15] W. Sorgel, W. Wiesbeck, EURASIP J. Appl. Signal Process. 2005 (2005) 296–305, http://dx.doi.org/10.1155/ASP.2005.296.
[16] A. Shlivinski, E. Heyman, R. Kastner, IEEE Trans. Antennas Propag. 45 (1997) 1140–1149, http://dx.doi.org/10.1109/8.596907.
[17] C. Baum, IEEE Trans. Electromagn. Compat. 44 (2002) 18–24, http://dx.doi.org/10.1109/15.990707.
[18] P. Miocinovic et al., Phys. Rev. D 74 (2006) 043002, http://dx.doi.org/10.1103/PhysRevD.74.043002.
[19] J.C. Hanson, L. Gerhardt, S. Klein, T. Stezelberger, S.W. Barwick, K. Dookayka, R. Nichol, Nucl. Instrum. Methods A 624 (2010) 85–91, http://dx.doi.org/10.1016/j.nima.2010.09.032.
[20] T. Barrella, D. Saltzberg, S.W. Barwick, J. Glaciol. 57 (2011) 61–66, http://dx.doi.org/10.3189/002214311795306691.
[21] Gerald Burke, Andrew Poggio, <http://en.wikipedia.org/wiki/Numerical_Electromagnetics_Code>.
[22] C. Balanis, Antenna Theory: Analysis and Design, 3rd ed., John Wiley and Sons, 2005.
[23] J. McClean, R. Sutton, Antennas and Propagation Society Intl. Symposium 3, 2003, pp. 571–574. doi:http://dx.doi.org/10.1109/ICUWB.2009.5288811.
[24] J. McClean, H. Foltz, R. Sutton, Conference on Ultra-Wide Band Systems and Technologies, 2004, pp. 317–321. doi:http://dx.doi.org/10.1109/UWBST.2004.1320987.
[25] S.A. Kleinfelder et al., The ARIANNA Collaboration, IEEE Trans. Nucl. Sci. 60 (2013) 612–618, http://dx.doi.org/10.1109/TNS.2013.2252365.
[26] C.S. Neal, Ann. Glaciol. 3 (1982) 216–221.
[27] J. Tatar, Performance of sub-array of arianna detector stations during first year of operation (Ph.D. thesis), University of California at Irvine, 2013.
[28] S.O. Rice, Bell Syst. Tech. J. 23 (1994) 282–332, http://dx.doi.org/10.1002/j.1538-7305.1944.tb00874.x.

[29] V.V. Bogorodsky, C.R. Bentley, P.E. Gudmandsen, Radioglaciology, D. Reidel Publishing, 1985.
[30] J. Kraus, R. Marhefka, Antennas, 3rd ed., McGraw-Hill, 2003.
[31] D. Seckel, Int. J. Mod. Phys. A 21-1 (2006) 70–74, http://dx.doi.org/10.1142/S0217751X06033398.
[32] P.W. Gorham et al., The ANITA Collaboration, Phys. Rev. Lett. 99 (2007) 171101, http://dx.doi.org/10.1103/PhysRevLett.99.171101.
[33] D. Saltzberg et al., Phys. Rev. Lett. 86 (2001) 2802, http://dx.doi.org/10.1103/PhysRevLett.86.2802.
[34] E. Zas, F. Halzen, T. Stanev, Phys. Rev. D 45 (1992) 362–376.
[35] L. Gerhardt, S. Klein, Phys. Rev. D 82 (2010) 074017, http://dx.doi.org/10.1103/PhysRevD.82.074017.
[36] K. Kotera, D. Allard, A.V. Olinto, JCAP 10 (2010), http://dx.doi.org/10.1088/1475-7516/2010/10/013.
[37] J. Alvarez-Muniz et al., Phys. Rev. D 62 (2000) 063001, http://dx.doi.org/10.1103/PhysRevD.62.063001.
[38] D. Heck, J. Knapp, J.N. Capdevielle, G. Schatz, T. Thouw, Report FZKA 6019, 1998, Forschungszentrum Karlsruhe.
[39] T. Huege, M. Ludwig, C.W. James. arXiv:1301.2132.